\pdfoutput=1

\documentclass[manuscript,screen,review=false]{acmart}

\usepackage{amsmath,amsfonts,amsthm} 
\usepackage{algorithmic}
\usepackage{textcomp}
\usepackage{graphicx}
\usepackage{fancyhdr}
\usepackage{soul}
\usepackage[utf8]{inputenc}

\usepackage{listings}

\usepackage{utfsym}

\usepackage{algorithm}
\renewcommand{\algorithmiccomment}[1]{\bgroup\hfill$\triangleright$~#1\egroup}

\usepackage{float}
\usepackage{CJK}
\usepackage{bm}

\usepackage{multirow} 
\usepackage{colortbl}
\usepackage{xcolor}
\definecolor{yellow}{rgb}{0.95,0.95,0.53} 
\newcolumntype{a}{>{\columncolor{yellow}}c}

\def \ie{i.e., }
\def \eg{e.g., }
\def \fig{Fig. }

\def \sec{Sec. }

\AtBeginDocument{%
  }

\setcopyright{acmcopyright}
\copyrightyear{2023}
\acmYear{2023}
\acmDOI{XXXXXXX.XXXXXXX}

\usepackage[colorinlistoftodos]{todonotes}

\begin{document}

\title{\textsc{Flip}: Data-Centric Edge CGRA Accelerator}


\author{Dan Wu}
\affiliation{%
  \institution{National University of Singapore}
  \city{Singapore}
  \country{Singapore}}
\email{danwu20@comp.nus.edu.sh}

\author{Peng Chen}
\affiliation{%
  \institution{Chongqing University of Posts and Telecommunications}
  \city{Chongqing}
  \country{China}
}
\email{chenpeng@cqupt.edu.cn}

\author{Thilini Kaushalya Bandara}
\affiliation{%
\institution{National University of Singapore}
  \city{Singapore}
  \country{Singapore}
  }
\email{thilini@comp.nus.edu.sg}

\author{Zhaoying Li}
\affiliation{%
  \institution{National University of Singapore}
  \city{Singapore}
  \country{Singapore}
  }
\email{zhaoying@comp.nus.edu.sg}

\author{Tulika Mitra}
\affiliation{%
  \institution{National University of Singapore}
  \city{Singapore}
  \country{Singapore}
  }
\email{tulika@comp.nus.edu.sg}

\renewcommand{\shortauthors}{Wu et al.}

\begin{abstract}
Coarse-Grained Reconfigurable Arrays (CGRA) are promising edge accelerators due to the outstanding balance in flexibility, performance, and energy efficiency. Classic CGRAs statically map compute operations onto the processing elements (PE) and route the data dependencies among the operations through the Network-on-Chip. 
However, CGRAs are designed for fine-grained static instruction-level parallelism and struggle to accelerate applications with dynamic and irregular data-level parallelism, such as graph processing. To address this limitation, we present \textsc{Flip}, a novel accelerator that enhances traditional CGRA architectures to boost the performance of graph applications. \textsc{Flip} retains the classic CGRA execution model while introducing a special data-centric mode for efficient graph processing.
Specifically, it exploits the natural data parallelism of graph algorithms by mapping graph vertices onto processing elements (PEs) rather than the operations, and supporting dynamic routing of temporary data according to the runtime evolution of the graph frontier. 
Experimental results demonstrate that \textsc{Flip} achieves up to 36$\times$ speedup with merely 19\% more area compared to classic CGRAs. Compared to state-of-the-art large-scale graph processors, \textsc{Flip} has similar energy efficiency and 2.2$\times$ better area efficiency at a much-reduced power/area budget. 
\end{abstract}

\ccsdesc[500]{Computer systems organization~Embedded systems}

\keywords{coarse-grained reconfigurable array, graph processing, accelerator}


\maketitle

Coarse-Grained Reconfigurable Arrays (CGRAs) are a prominent approach among high-performance low-power edge accelerators due to {their} excellent balance between flexibility, performance, and energy efficiency. 
Existing CGRAs
employ a static \textit{operation-centric} approach in which the operations within the dataflow graph (DFG) of the loop kernel are spatially mapped onto the processing element (PE) array, and the data dependencies among the operations are routed through on-chip network (NoC) \cite{podobas2020survey, martin2022twenty, li2022coarse, liu2019survey}.
 In other words, CGRAs rely on operation-level parallelism to accelerate loop kernels. Furthermore, the placement of operations and data routing paths are statically determined at compile time. 
 The classic CGRA execution model exhibits the following key characteristics: (1) The PE array exploits mainly operation-level parallelism and does not utilize data-level parallelism if there are data dependencies in the input data; (2) All data are stored in the scratchpad memory~(SPM), and each iteration necessitates regular SPM access; (3) The static nature of the fine-grained operation mapping requires the majority of instruction-level parallelism to be known at compile time to facilitate pipelining. \cite{riptide, voitsechov2014single}. 
 Thus, CGRAs are suitable for kernels with high operation-level parallelism and low-intensity, regular memory access.

However, CGRAs are not well-suited for applications with irregular memory access, poor data locality, runtime control, and data dependencies that do not expose parallelism at compile time and inherently conflict with static mapping. Accelerating these irregular applications is crucial for the broader adoption of CGRA technology. Graph processing, a vital application at the edge (e.g., to support indoor or outdoor navigation), features intensive, irregular memory access and light computation with minimal instruction-level parallelism ~\cite{gui2019survey, mccune2015thinking}. Consequently, graph applications exhibit poor performance on CGRAs. Moreover, while CGRAs can exploit fine-grained instruction-level parallelism, static mapping makes it challenging to accelerate applications with inherent coarse-grained dynamic data parallelism. 
Graph algorithms typically exhibit natural parallelism across vertices in the dynamically evolving graph frontier, i.e., vertices within the frontier can be processed in parallel. 
To harness this coarse-grained parallelism, graph accelerators iteratively dispatch tasks (e.g., active vertices) to different processing units, synchronize, and update the task queue (e.g., graph frontier). This complex schedule and dispatch mechanism is suitable for large graph processing accelerators~\cite{nguyen2021fifer, dadu2021polygraph, rahman2020graphpulse, zhang2021depgraph, stevens2021gnnerator}
with significant area and power budgets. Although graph processing is widely used in edge devices, such as pathfinding in network devices~\cite{guthaus2001mibench} and navigation in small robots, tiny graph accelerators that can operate within limited power budgets are lacking. 

We present \textsc{Flip}, a tiny CGRA accelerator capable of supporting both the traditional {\em operation-centric} execution model for compute-heavy regular kernels and a novel \textit{data-centric} paradigm that harnesses the natural coarse-grained parallelism of the graph algorithms. The data-centric paradigm employs a vertex-centric programming model, wherein the compute function of a vertex receives the updated vertex value from its neighbor, calculates a new vertex value, and distributes the updated value to neighboring vertices~\cite{mccune2015thinking}. 
\textsc{Flip} upends the operation-centric paradigm by distributing data graph vertices to the PEs and routing the edges rather than mapping the operations corresponding to the loop kernel. 
Indeed, as the vertex processing code is typically quite short and the same for all the vertices, 
this code is stored in each PE and executed sequentially for a vertex. Once a vertex's processing is completed, the current PE sends the intermediate result to PEs where the neighboring vertices are mapped and triggers their processing in parallel. The execution thus follows the parallelism unearthed by the frontier of active vertices instead of struggling to find the meager parallelism among the operations within the computation for a single vertex. 

\begin{figure}[t]
    \centering
     \includegraphics[trim=0cm 0cm 0 0cm, clip, width=3.4in]{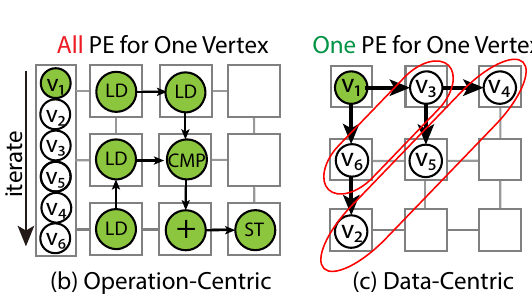}
    \caption{Operation-centric vs. Data-centric execution modes. Operation-centric mode maps the operations on the PEs and processes the vertices stored in scratchpad memory iteratively. Our proposed data-centric mode maps the vertices on different PEs and processes them in parallel with identical code. }
    \label{figure:two_modes}
\end{figure}

\fig \ref{figure:two_modes} illustrates the differences between the operation-centric and data-centric execution models for the Breadth-First Search (BFS) algorithm. 
The data graph to be processed is shown in \fig \ref{figure:two_modes}(a).  Assuming BFS is invoked with $V_1$ as the source vertex, \fig \ref{figure:two_modes}(b) displays the traditional operation-centric execution model in CGRA. In this model, operations to process one edge of a vertex are spatially mapped on the PE array along with the routing of the data dependencies among the operations. 
Since concurrent processing of multiple vertices can potentially cause data conflict due to non-atomic read/write pairs, 
this approach can only process one vertex at a time but takes advantage of parallelism among the operations, which is quite limited in this scenario. 
 In contrast, the data-centric execution model of \textsc{Flip}, shown in \fig \ref{figure:two_modes}(c), statically maps data vertices onto the PE array. Execution is triggered by the PE containing the source vertex $V_1$ where the code for processing the vertex is executed sequentially.
The completion of processing $V_1$ triggers parallel processing of $V_3$ and $V_6$, and their completion, in turn, triggers parallel processing of the new frontier~($V_2$, $V_4$, $V_5$). Consequently, data-centric execution exploits coarse-grained dynamic data parallelism. 

Achieving data-centric execution model on low-power CGRA accelerator is not straightforward. Different graph algorithms and even invocations of the same algorithm with different starting points lead to different parallelism and data routing patterns. 
In \textsc{Flip}, we augment the underlying classic CGRA architecture to support dynamic routing of the intermediate data along the graph edges (e.g., $V_1$ to $V_3$ and $V_6$). We also design a compiler that can map the vertices onto the PE array to expose parallelism among the frontier vertices (e.g., $V_2$, $V_4$, $V_5$ should preferably be mapped to distinct PEs) while ensuring short routing lengths for the edges. 
Finally, for larger graphs that cannot fit into the on-chip memory, we partition the graph and swap partitions in and out at runtime. 
To support dynamic graphs with changing vertex or edge attributes, we update these attributes while the corresponding partition is swapped out to the main memory, utilizing a general-purpose processor (e.g., micro-controller unit) for the updates.
    
\begin{table}[t]\footnotesize
  \centering
  \caption{Comparison among CGRA-based accelerators.}
  \begin{tabular}{c c c c c a}
  \hline
      	& \textbf{PolyGraph}  & \textbf{Fifer}  & \textbf{HyCUBE} & \textbf{RipTide} &  \textbf{\textsc{Flip}}\\

      	& \cite{dadu2021polygraph} 	  	& \cite{nguyen2021fifer} 		&  \cite{wang2019hycube}   & \cite{riptide} 	& (\textit{this work})\\
 \hline
Graph Application Performance	&
      $\usym{2714}$ & $\usym{2714}$ & $\usym{2717}$ & $\usym{2717}$ & $\usym{2714}$\\
General Application Performance
    	& $\usym{2717}$ & $\usym{2714}$ & $\usym{2714}$ & $\usym{2717}$ & $\usym{2714}$\\
     Power Efficient &  $\usym{2717}$ &  $\usym{2717}$  & $\usym{2714}$ & $\usym{2714}$ & $\usym{2714}$ \\
     Area  Efficient &  $\usym{2717}$	&  $\usym{2717}$ &  $\usym{2714}$  & $\usym{2714}$ & $\usym{2714}$ \\ \hline
Number of PEs	& 16$\times$5$\times$4  & 16$\times$16$\times$5 & 4$\times$4  & 6$\times$6  & 8$\times$8 	\\
    CGRA mode 	& Op-Centric & Op-Centric & Op-Centric & Op-Centric & \textbf{Data-Centric}\textbf{\&Op-Centric}\\
\hline
  \end{tabular}
  \label{table:sota_CGRA_summary}
\end{table}

To the best of our knowledge, \textsc{Flip} is the only programmable  accelerator under a limited power budget that supports graph processing at the edge while also accelerating traditional compute-intensive kernels that are a natural fit for CGRA architecture. 
Table \ref{table:sota_CGRA_summary} presents a qualitative comparison of \textsc{Flip} with recent CGRA-based accelerators. 
Data-center scale advanced graph accelerators, PolyGraph~\cite{dadu2021polygraph} and Fifer~\cite{nguyen2021fifer}, adopt multi-core graph processor design with complex components like scheduler, control core, caches, and a complete CGRA fabric within each core. In these hierarchical architectures, the CGRA serves only as a low-power substitute for the compute core, offering instruction-level parallelism.
 The higher-level core array captures coarse-grained data parallelism, thanks to the complex control components.
In the low-power domain, such a complex system design is infeasible due to power constraints. With a single and simple CGRA fabric, HyCUBE~\cite{wang2019hycube} and RipTide~\cite{riptide} adopt the operation-centric paradigm to support diverse applications and thus perform poorly on graph applications. 

Experimental evaluation shows that: (1) The proposed data-centric model can achieve high data-level parallelism compared to the conventional operation-centric approach. 
(2) On graph processing tasks, \textsc{Flip} achieves up to 36$\times$ performance speedup with only a 19\% increase in area compared to the operation-centric CGRA accelerator.
(3) Compared with an advanced graph processor, PolyGraph\cite{dadu2021polygraph}, 
\textsc{Flip} achieves similar performance per watt and 2.2$\times$ performance per area at only 1.1\% and 0.5\% area and power budget, respectively.




\section{Background and Motivation}

\subsection{ Graph Processing at the Edge}
Unlike large-scale graphs, graphs at the edge exhibit distinct characteristics. We leverage these unique features of graph processing at the edge, as opposed to processing large graph data in the cloud, to achieve data-centric CGRA acceleration at the edge.
\\
\textit{Graph algorithms can be expressed through programming models }\cite{gui2019survey}, such as the vertex-centric programming model, where the same concise function is executed for each vertex in the graph \cite{besta2019graph}. In other words, vertex data stored on different PEs can be processed with the same small code snippet, requiring minimal instruction memory per PE. The concept is akin to the SPMD (single program, multiple data) computing model employed by GPUs, except for the rich and dynamic dependencies among the vertices, contrasting with the massive regular data parallelism in most GPU kernels. 
\\
\textit{Graph processing at the edge typically employs small graphs}, such as neighborhood road network for navigation \cite{guthaus2001mibench}.
In these cases, graph data can be stored entirely on-chip. The graphs usually have limited in- and out-degree, resulting in a small and balanced memory requirement for maintaining the edge data corresponding to each vertex. With only on-chip distributed memory inside the PE array, \textsc{Flip} can store $256$ vertices, roughly 2.5 $km^2$ downtown in San Francisco\cite{li2005trip}. With a centralized memory unit, like the main memory of a micro-controller unit or DRAM, \textsc{Flip} can also handle larger problem sizes through data swapping from the centralized memory (\sec \ref{section:data_swap})
\\
\textit{Most graph algorithms maintain a static graph structure, modifying only the attributes of vertices and edges.} 
In this scenario, the routing configurations for most edges remain static. So we can map a graph once and run multiple queries on the same data, or even run multiple applications on it by only replacing the code snippet on PEs. \textsc{Flip} also supports efficient attribute changing for dynamic graphs. Without recompilation, or bulky data movement, only the graph partitions with the changed vertices/edges are swapped out and updated.


\begin{figure*}[ht]
\centering
\includegraphics[width=\textwidth]{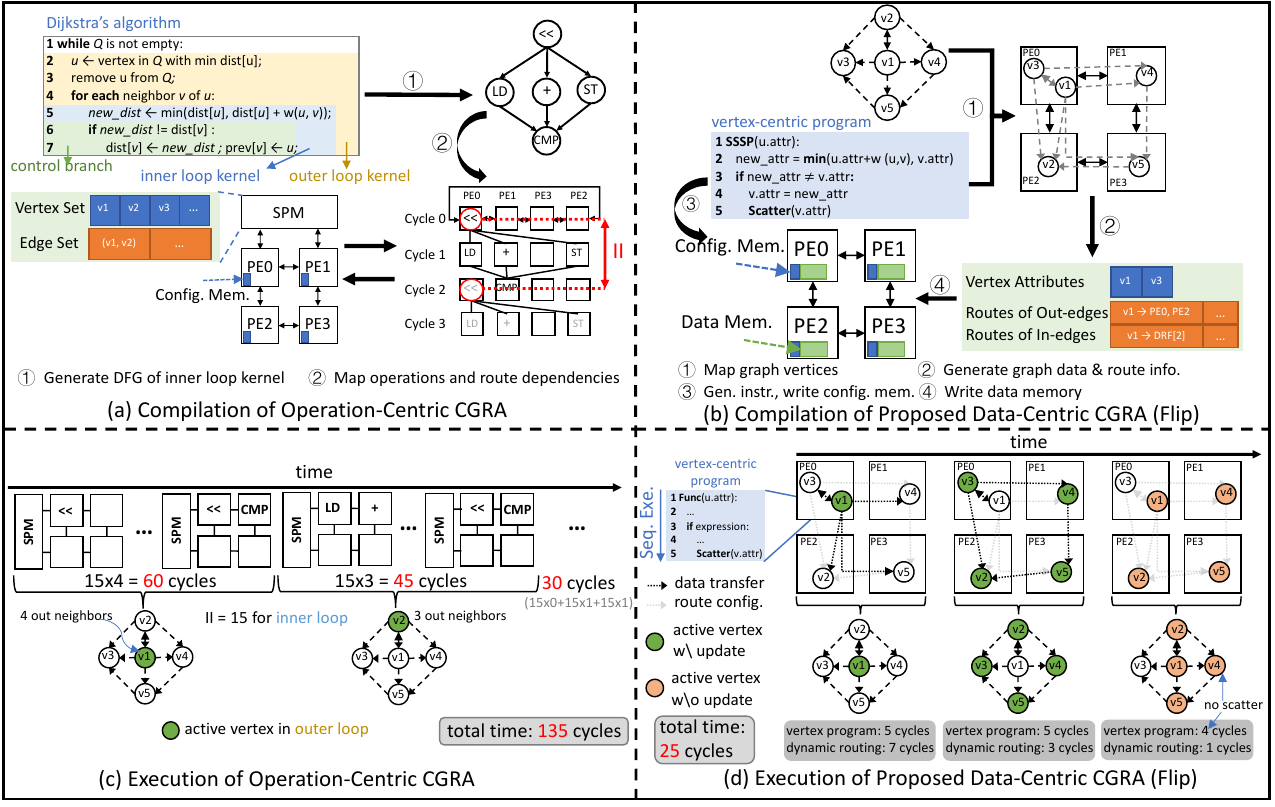}
\caption[overview]{Compilation and execution for operation-centric CGRA and data-centric CGRA, taking SSSP problem as an example. Data-centric CGRA benefits from data-level parallelism and outperforms operation-centric CGRA which suffers from low instruction-level parallelism in graph algorithms.}
\label{figure:data_centric_model}
\end{figure*}

\begin{figure}[]
    \scriptsize
	\centering
\includegraphics[width=3.4in]{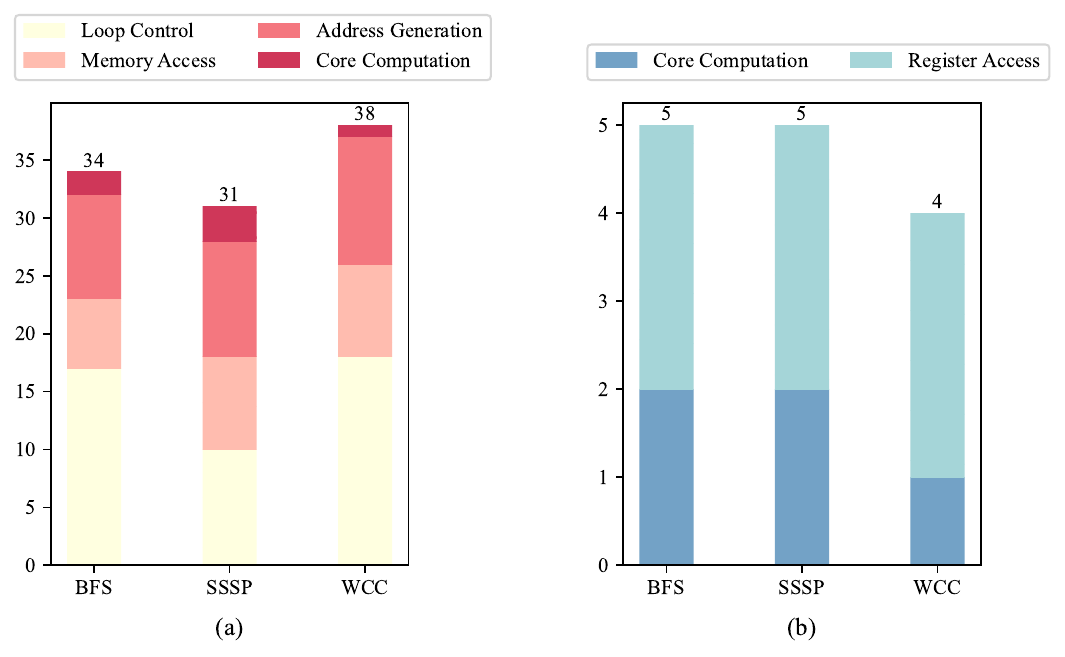}
 \caption{Number of operations for (a) operation-centric and (b) data-centric CGRA. \textit{Memory Access} counts only graph data access.}
	\label{figure:motivation_ratio_ops}
\end{figure}

\begin{figure}[t!]
   \centering
    \includegraphics[width=4in]{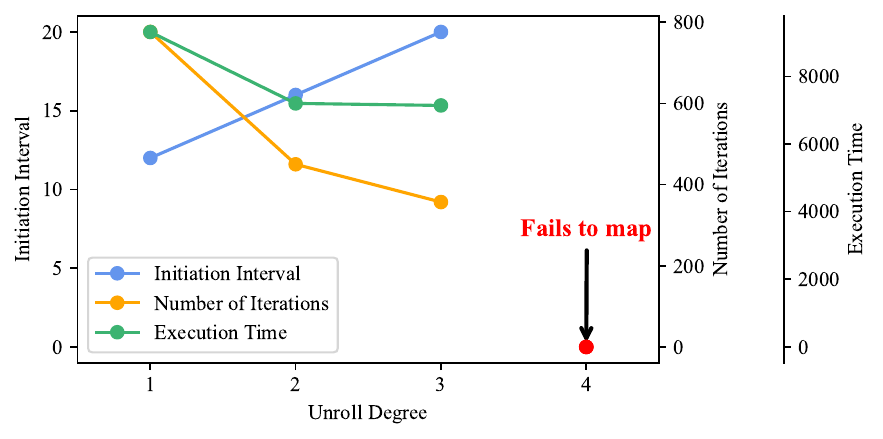}
    \vspace{-10pt}
    \caption[unroll]{Speed up with different degrees of unrolling on operation-centric CGRA for BFS on road networks. The execution time smoothes when the unroll degree reaches three and achieves only 1.3x speedup. Because of the exponentially growing compilation time, further unrolling will result in compilation failure. }
    \label{figure:unroll}

\end{figure}


\subsection{Motivating Example}

Conventional operation-centric CGRAs accelerate application execution through instruction-level parallelism. Consider the example depicted in Fig. \ref{figure:data_centric_model}(a), where the inner loop kernel (line 5-7) of Dijkstra's algorithm for SSSP is mapped onto a 2x2 conventional CGRA. The compiler first converts the loop kernel to DFG representation (\textcircled{\raisebox{-0.9pt}{1}}). A toy DFG is used for illustration. Then it schedules the DFG onto the time-extended 
routing resource graph in a modulo fashion (\textcircled{\raisebox{-0.9pt}{2}}). The schedule repeats every two cycles, which is the Initiation Interval (II). This modulo schedule is loaded into the PE local configuration memory, and the graph data is loaded to the SPMs before execution.
We can find the detailed execution in \fig \ref{figure:data_centric_model}(c). Each execution of the loop iteration needs to load (store) data from (to) SPM. For the sake of clarity and without loss of generality, in this example, we assume the edge weight to be 1, which means that each edge needs to be processed only once.

The operation-centric model has several limitations for graph processing. The intensive and irregular memory accesses demand considerable operations for address calculation and data fetching.
As shown in \fig~\ref{figure:motivation_ratio_ops}(a), 20\% of the DFG operations are used to access graph data and 30\% for memory address generation. Furthermore, the irregular memory access leads to substantial memory bank conflicts, further limiting memory-level parallelism. Additionally, without complicated hardware support, static scheduled CGRAs cannot support parallel graph applications \cite{buluc2017distributedmemory, 10.1145/3466752.3480133, 7234934} designed for multi-core processors with coarser parallelism (per-core graph partition) due to the cost of dynamic communications and synchronization. Concurrently, fine-grained parallelization methods like loop unrolling struggle to provide the expected speedup due to the dependencies among the vertices and varying number of neighbors, as shown in \fig \ref{figure:unroll}. The increased size of DFGs after unrolling also leads to compiler crashes due to exponentially increasing complexity of mapping problem.
Lastly, loop control operations identifying neighboring vertices constitute a substantial fraction of the total operations, as show in \fig~\ref{figure:motivation_ratio_ops}(a).
Overall, it takes 15$\times$9=135 cycles to process all nine edges in this example, and only one vertex can be processed at a time.

\textsc{Flip} addresses all these issues through data-centric paradigm. As shown in \fig \ref{figure:data_centric_model}(b), instead of utilizing the entire PE array for inner loop computation, corresponding to processing one edge, we directly map the graph to the PE array. Each PE stores the vertex attributes mapped to it, attributes of the incoming edges, and the destination PEs of outgoing edges. By mapping data to PEs, the number of instructions is minimized, as shown in \fig \ref{figure:motivation_ratio_ops}(b). Since data needed by one PE is either stored locally or passed from other PEs through the NoC, address calculation is no longer required. Moreover, each PE executes the vertex-centric program independently, so instructions for loop control are eliminated. For larger graphs, different parts (multiple vertices) can be mapped onto the same PE through time-multiplexing and swapped in and out at runtime.

A PE executes the code snippet (5 cycles for SSSP) sequentially when triggered by an incoming edge, then route the data (updated vertex's attribute) through the NoC.
In the data-centric model, an application starts at a designated vertex (e.g., source $V_1$ in SSSP) which is not known at compile time. The intermediate results (e.g., distance from $V_1$ to the current vertex) are sent to the neighboring vertices, which are then activated. In \fig \ref{figure:data_centric_model}(d), the neighbors of vertex $V_1$ ($V_2$, $V_3$, $V_4$, $V_5$) are mapped to different PEs, enabling parallel processing. A PE only needs information from the incoming packet and the local data during the processing. In this example, it takes 5+7=12 cycles to process $V_1$ and scatter data to its neighbors, then another 7 cycles to process and scatter $V_2$, $V_3$, $V_4$, $V_5$ in parallel. Finally, vertex $V_1$, $V_2$, $V_4$ and $V_5$ are activated but the program stops early in 4 cycles as there is no attribute update and thus nothing to scatter. It takes a total of 25 cycles to process the 5 vertices, as opposed to 135 cycles in the operation-centric model. 


\section{\textsc{Flip} Overview}
\begin{figure}[t]
    \centering
    \includegraphics[width=0.3\textwidth]{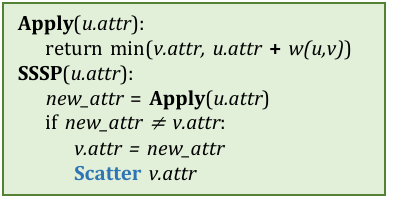}
    \caption{Implementation of SSSP in the vertex-centric programming model. Scatter instruction triggers the message passing to destination vertices.}
    \label{figure:program_example}
    \end{figure}

We design \textsc{Flip} to enable efficient graph processing on CGRA accelerators. {\em Note that the underlying architecture supports operation-centric execution mode, which is not discussed as it is similar to classic CGRA.} We detail the data-centric mode for graph processing. \textsc{Flip} uses the vertex-centric programming model where a vertex is processed following the general sequence of receiving updated value from its neighbor, calculating new values (\textit{Apply()}), and distributing the updated values to neighboring vertices (\textit{Scatter})\cite{mccune2015thinking}. 
    \fig \ref{figure:program_example} shows the example code of single source shortest path (SSSP) written in this model. 
\textsc{Flip} takes the graph as input and maps the graph vertices onto the PE array in a many-to-one manner. Similar to some graph processors \cite{yao2022scalagraph,rahman2020graphpulse,ahn2015scalable}, \textsc{Flip} follows a graph frontier approach with the vertex-centric programming model.
The key difference is that prior graph processors use programs to schedule and control the computation in each processing unit with data fetched from memory at runtime, while \textsc{Flip} statically places the data onto the PEs and uses the frontier data to trigger processing in each PE directly. This reduces control overhead tremendously and enables asynchronous distributed graph processing.
The graph application execution is triggered from a specific vertex on a PE (\eg the current location in navigation systems) and progresses to other vertices.

We also design an efficient compiler to map the graph data onto the PE array. The compiler combines locality and parallelism to maximize graph processing performance.
The compiler generates the routing information for graph edges. The graph is mapped offline. Before execution, the generated configuration, containing vertex properties, instructions, and routing information, is distributed to the PE array from the host. User specifies the start vertex during execution. \textsc{Flip} supports graph attribute update through data swap during execution. 


\section{\textsc{Flip} Architecture}
\label{section:architecture}

\begin{figure}[t]
\centering
\includegraphics[width=0.7\textwidth]{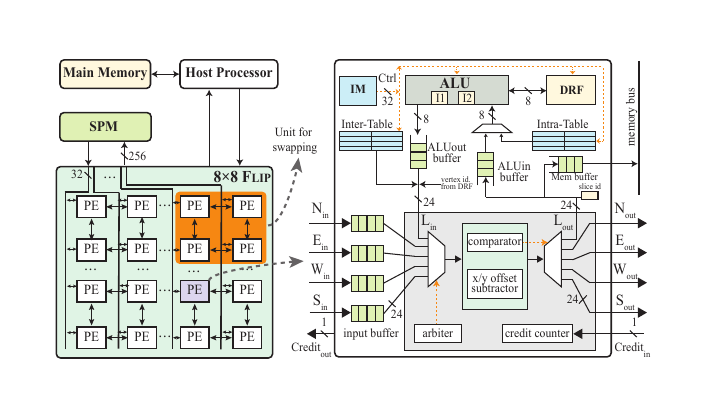}
\caption{\textsc{Flip}: CGRA enhanced with  \textit{data-centric} mode. The arbiter, comparator and x/y offset subtractor are used for dynamic routing. Once a packet reaches the destination PE, the Intra-Table is used to find the vertex location in Data Register File (DRF) and the edge weight. To scatter a vertex's new attribute, the value is pushed to the ALUout buffer and will be latter packed with the x/y offset of destination PE(s) in Inter-Table.}
\label{figure:overview}
\end{figure}

\fig \ref{figure:overview} shows the proposed \textsc{Flip} architecture. \textsc{Flip} fabric contains a PE array (8$\times$8 in our prototype) connected by a mesh network. The complete \textsc{Flip} system also contains on-chip SPM (16KB in 8 banks),  off-chip memory (256KB), and a scalar core as the host processor.
To support data-centric mode, \textsc{Flip} uses on-chip cross-PE distributed memory (64$\times$260B = 16KB) for graph data storage and packet buffering, and a NoC supporting dynamic routing. 

Each cycle, an incoming packet is selected by the arbiter and then routed by offset subtractor and comparator. After a packet reaches the destination PE, the carried slice id is compared with the stored slice id in PE. If they are the same, the content carried by the packet will be later used for computation. Otherwise, this packet will be pushed to memory buffer and sent to SPM. For an updated vertex to be scattered, the updated vertex attribute is pushed to the ALUout buffer, packed with the inter-PE routing information in Inter-Table. Finally the packet is pushed to the NoC.
 
For large problem sizes, \textsc{Flip} supports efficient data swapping at runtime.
To take advantage of data locality while avoiding straggler problems, we use non-overlapping 2$\times$2 PE clusters as the basic units of data swapping. The graph partition mapped onto a PE cluster is called a \textit{slice}. Multiple slices can be mapped onto the same PE cluster, identified by a unique id $slice_{id}$.
We use the same ISA as classic CGRAs, ensuring that the underlying CGRA can be used for static mapping of non-graph applications by deactivating dynamic routing.

\subsection{PE Memory}
\label{sec:Memory}
Distinct from conventional CGRAs, \textsc{Flip} has larger memory per PE to handle contention at runtime and hold the data originally stored in the multi-bank SPM. One PE in \textsc{Flip} consists of seven storage components for different functionalities. All buffers in \textsc{Flip} are FIFO queues.\\
\textbf{Instruction Memory (IM)} is a register file holding the instruction set of the vertex-centric program. The code snippet is identical for all PEs.\\
\textbf{Data Register File (DRF)} stores the properties (\eg vertex index, BFS level) of the vertices that are mapped onto the current PE.
The properties of the vertices can be updated during execution.
With 4 registers per DRF, we allow at most 4 graph vertices to be stored in each PE.\\
\textbf{Slice ID Register} is an 8-bit register storing the $slice_{id}$ of currently loaded subgraph (slice).\\
\textbf{ALUout Buffer} 
\label{section:aluoutbuffer} stores packets in the format  ($\textit{id}_u$, $\textit{attribute}_u$), where $u$ is the current vertex with an updated attribute and needs to send the value to its neighboring vertices for update.
When the packet is sent out to the adjacent PE, we add an $\textit{offset}_v$ item indicating the number of hops in x and y axis to the PE that stores the destination vertex $v$.\\
\textbf{Input Buffers}
\label{section:inbuffer} cache the incoming packets from adjacent PEs. We have one input buffer per input port. 
Without contention-free routing, we use larger input buffers to store multiple packets and avoid potential deadlock or packet loss.
The packets in the input buffer have the format  ($\textit{id}_u$, $\textit{offset}_v$, $\textit{attribute}_u$, $slice_{id}^v$).\\
\textbf{ALUin Buffer} \label{section:aluinbuffer} holds the packets forwarded by the routing logic till ALU is ready to receive new packets. Each incoming packet is processed and updated with edge attributes before being fed to ALU.
\\
\textbf{Memory Buffer} \label{section:memorybuffer} caches packets whose destination vertex is currently off-chip. To check whether the destination vertex $v$ is loaded, the $slice_{id}^v$ inside the packet is compared with the value in Slice ID Register. Then the packet will be pushed into either ALUin buffer or Memory buffer according to the comparison.
\\
\textbf{Inter-Table} stores the \textit{inter-PE} routing information, identifying the PEs that house the destination vertices for each vertex within this PE (Sec.~\ref{sec:inter-table}).
\\
\textbf{Intra-Table}, on the other hand, maintains the \textit{intra-PE} routing information and incorporates edge weights, specifying the target register(s) storing the destination vertex (vertices) for packets destined for this PE as well the weight for this edge (Sec.~\ref{sec:intra-table}).

\subsection{Dynamic and Contention-Aware Routing}
\label{section:route}
The graphs we are processing at the edge have limited in-/out-degree, so the PEs have a balanced communication workload. 
Thus, there is limited benefit
in supporting routes with an arbitrary number of turns that are useful to bypass the busy area.
Therefore, we only support YX dimension-ordered routing \cite{dally2004principles, jerger2017chip}, which requires limited routing configuration and saves memory cost. 
We set two routing tables in each PE, storing the graph edge and routing information compactly, as shown in \fig \ref{figure:table}. 
With these two routing tables, a PE can dynamically select the corresponding routing configuration via finding offsets to destination PEs and identifying the address of the destination vertex inside the PE.
The addressing process 
is thus split into two stages: inter-PE and intra-PE addressing.
Inter-PE addressing aims to find the target PE where the neighboring vertices locate, while intra-PE addressing is to find the specific vertex among multiple vertices inside current PE.

\subsubsection{Inter-PE Addressing}
\label{sec:inter-table}
\begin{figure}[t]
\centering
\includegraphics[width=0.7\textwidth]{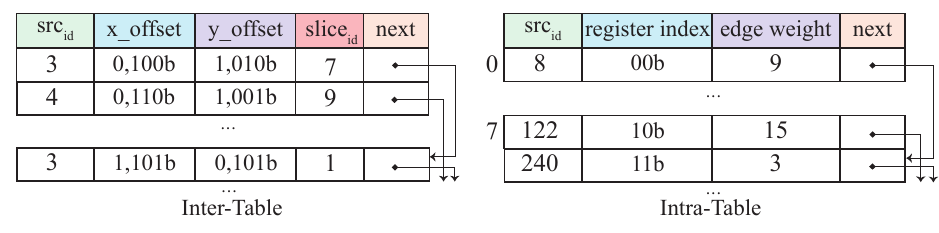}
\caption{Example of routing table in \textsc{Flip}.}
\label{figure:table}
\end{figure}

An Inter-Table is established inside each PE for inter-PE addressing.
As shown in \fig \ref{figure:table}, each Inter-Table entry indicates the offsets to the target PE, consisting of source vertex index $\textit{src}_{id}$, $x\_\textit{offset}$ in the x dimension, $y\_\textit{offset}$ in the y dimension  and $slice_{id}$ of the slice containing destination vertex.
 $x\_\textit{offset}$ and $y\_\textit{offset}$ have 4 bits each, of which the first bit indicates the transmission direction (\ie 0 for negative direction, and 1 for positive direction), and the subsequent 3 bits (i.e., up to 7 hops) refer to the number of hops to the destination along that dimension.
For example, in Inter-Table of \fig \ref{figure:table}, the first source vertex index is 3 and the corresponding $x\_\textit{offset} = \{0,100\}$ and $y\_\textit{offset} = \{1,010\}$.
This means that the packet generated by vertex 3 needs to be sent out first along the $+y$ dimension with 2 (=$0b010$) hops, then along the $-x$ dimension with 4 (=$0b100$) hops.
Thus, the packet needs to carry the routing information of $x\_\textit{offset}$ and $y\_\textit{offset}$ values to find the target PE.
The corresponding offset is decremented by one after each hop.
The packet is delivered to the destination PE when the last three bits of $x\_\textit{offset}$ and $y\_\textit{offset}$ become zero.

To accelerate the search within Inter-Table, we organize the table entries with a linked list.
All the entries with the same source vertex index (e.g., 3) are organized as a list;
there is a head entry that can be found sequentially from the table head, which points to the second entry with the same vertex index, and the pointer field (\ie next) of its last entry is \texttt{NULL}.
The list length is equal to the outgoing degree of the current vertex, which is short since \textsc{Flip} is proposed for low-degree graphs.
The four head entries, one for each vertex of current PE, with the unique source vertex index, are placed at the four headmost positions of Inter-Table for fast search.

\subsubsection{Intra-PE Addressing}
\label{sec:intra-table}
When a packet reaches the target PE and its $slice_{id}$ matches with the value in Slice ID Register, intra-PE addressing helps find the specific vertex among the multiple vertices inside the target PE.
An Intra-Table entry indicates destination vertex and incoming edge attributes, consisting of the index $\textit{src}_{id}$ of source vertex of the incoming edge, register index for destination vertex in DRF of current PE, and edge weight for the current edge.
$\textit{src}_{id}$ of Intra-Table entry has 8 bits, used to identify the destination vertex and the corresponding edge weight.

To accelerate the search within Intra-Table, all the entries are organized as multiple linked lists similar to Inter-Table.
The source vertex index of entries in the same list has the same hash value returned by a hash function (i.e. $src_{id}$\%8).
Before table searching for incoming packets, the hash function is computed to find the head entry address of the corresponding list.
The 8 head entries (\ie hash value ranging from 0$\sim$7) are placed at the headmost position of Intra-Table.
The entries inside the list are searched sequentially.
In the experiment, the average length of each list is less than 2, and the average search time inside the list is no more than 2 cycles.




\subsubsection{Flow Control}
Distinct from the static routing in conventional CGRAs, the packets in \textsc{Flip} are dynamically generated, and there are potential contentions for communication resources (\eg output port) in the NoC. We adopt credit-based routing \cite{dally2004principles, jerger2017chip} to handle contention.

\subsection{Runtime Data Swapping}
\label{section:data_swap}
When the PE array cannot hold a graph due to capacity limit, the off-chip memory is used to store the whole graph, and graph partitions are brought to \textsc{Flip} at runtime using SPM as a cache. Data swapping can also be used to update the vertex/edge attributes if the graph structure remains unchanged, e.g., real-life traffic on road networks, through popping slices out and swapping them in after modifying the values. 

At runtime, if a packet cannot find its destination vertex on-chip, the packet is cached first, then loaded and executed after the corresponding vertex is brought in. Specifically, if the $slice_{id}$ does not match once a packet reaches the destination PE, the packet is pushed into \textit{memory buffer} and sent to SPM. 
Once a PE cluster is idle, data swapping is initiated. We adopt a simple cache-friendly priority; the slice with the earliest pending task (cached packet) is loaded among all the slices mapped to the same PE cluster.

\subsection{Mode Switching}
\label{section:two_mode}
\textsc{Flip} architecture natively supports operation-centric computing. In this mode,  Inter-Table and Intra-Table store the pre-determined Xbar configurations for static routing. Instructions are stored in \textit{Instruction Memory} (IM) similar to data-centric mode.
 A global trigger initiates execution of all the PEs, and a global program counter is used to cycle through the configurations in all PEs in sync as opposed to an independent program counter for each PE in data-centric mode. Components like comparator, offset subtractor etc. are disabled and the SPM is used for centralized data storage in operation-centric mode.
\section{\textsc{Flip} Compiler}
\label{section:compiler}
In this section, we discuss the mapping of graph data onto the \textsc{Flip} PE array in the data-centric mode. We first explore the mapping algorithm when the entire graph can fit into the PE array, and subsequently introduce how to handle larger problem sizes.

\subsection{Problem Formulation}
The \textsc{Flip} compiler maps the graph vertices to PEs.   Once the placement is ensured, a YX-dimension ordered route is generated for each graph edge.
Since the \textsc{Flip} compiler assumes an unknown starting vertex with non-deterministic runtime contention, the final run time cannot be derived during compilation. 
Therefore, we devise two mapping objectives that are closely correlated with the run time: the \textit{total routing length} and \textit{sequentialization}.\\
\textbf{Total Routing Length}: As vertex updates are transmitted along the edges, and each edge is mapped onto a series of links on NoC, the total routing length of all graph edges determines the communication cost. The longer the routing length, the higher the communication cost. Since one-hop routing latency is costly in our contention-tolerant  NoC (close to the computation time of one packet), routing length plays a predominant role in evaluating mapping quality.\\
\textbf{Sequentialization}:
Simply clustering vertices for locality can hurt parallelism among the frontier vertices.
Despite the dynamic routing and unknown starting vertex, there is a guaranteed barrier to parallelism: two vertices mapped to a PE have to be sequentially executed if they have incoming edges from the same vertex. Thus, a high-quality mapping should balance locality and parallelism.
\\
\textbf{Problem Definition:}
Given a graph $G(V,E)$ and a \textsc{Flip} instance with PE array $P$, the problem is to construct a many-to-one mapping $M = \{(v, p)\}$ of vertices in $G$ to PEs in $P$.
\begin{itemize}
    \item Each vertex $v \in V$ should be mapped to one PE $p \in P$.
    \item Each PE $p \in P$ can have at most $n$ graph vertices mapped onto it. In the current design, $n = 4$.
    \item Minimize \textit{total routing length} and \textit{sequentialization}.
\end{itemize}

\begin{algorithm}[h]
\caption{\textsc{Flip} Mapping Algorithm}
\label{alg:mapping}
\begin{algorithmic}[1]
\REQUIRE Graph $G(V,E)$ and PE array $P$  
\ENSURE Mapping $M=\{v:p\}$ for $v \in V, p \in P$ 

\STATE P = P.replicate($\lceil\frac{|V|}{P.capacity} \rceil$);
\STATE $M$ = \{$v_c$:$p_c$\}; 
\STATE $M$ = Beam\_Search($M$); \COMMENT{Generate initial mapping} 
\WHILE[Local optimization]{$M$ is not stable}
    \STATE $p$ = random(P); $P_p$ = Neighbors of $p$; 
    \STATE $V_p$ = vertices mapped on $p$; $V_P$ = vertices mapped on $P_p$; 
    \STATE $\psi$ = combination($V_p$, $V_P$); 
    \STATE best\_pair = $M$.\textbf{estimate}($\psi$, $t_{h}$);
     \STATE If best\_pair.cost$>$0, swap the placement of best\_pair in $M$; 
\ENDWHILE
\end{algorithmic}
\end{algorithm}

\subsection{Mapping Algorithm}
\label{section:mapping}
    Algorithm~\ref{alg:mapping} presents the pseudo-code of \textsc{Flip} mapping algorithm. It adopts a two-phase strategy to generate high-quality mappings. The first phase creates an initial mapping optimized only for the routing length with beam search~\cite{reddy1977speech}. The second phase optimizes the mapping locally to reduce sequentialization and improve parallelism, guided by a \textit{run time estimation model} (Algorithm~\ref{alg:estimation_model}).
    
   \subsubsection{Initial Mapping Generation}  \label{sec:beam_search}
    In the first step, a mapping is generated using beam search with beam width $k$=10. Beam search is a heuristic that explores the search tree by sorting the nodes at each level with a pre-defined function and only exploring $k$ most promising nodes, known as beam width. We only use the primary objective, \textit{total routing length}, in beam search, denoted as $f(M)$. For an incomplete mapping $M'$ where just a subset of $V$ is mapped, only edges with both ends mapped are evaluated in $f(M')$.\\
    Initially, the graph center $v_c$ with minimum eccentricity is placed at the center $p_c$ of the PE array. This incomplete mapping is the root of the search tree.
    Formally, we denote the root as 
    $S(0,M,V_{can},P_{can})$ containing: layer in the search tree which is equal to the number of vertices mapped minus 1, incomplete mapping $M$=$\{v_c$:$p_c\}$, candidate vertices $V_{can}$=\{neighbors of $v_c$\} to map for the next iteration, and candidate PEs $P_{can}$=\{$p_c$ and PEs next to $p_c$\} available  for mapping vertices in the next iteration.
    At level $i$ of the tree, it expands all nodes at the current level by generating their successors.
    A successor $S(i$+1$,M',{V'}_{can},{P'}_{can})$ of a node $S(i,M,V_{can},P_{can})$ is obtained by (a) assigning a vertex $v$ in $V_{can}$ to a PE $p$ in $P_{can}$, (b) updating the frontier-like candidate vertex set and candidate PE set. Only top $k$ nodes with the lowest $f(M)$ are kept in a layer of the search tree.

    \subsubsection{Local Optimization to Balance Locality and Parallelism}
    \label{section:local_optimization}
 As beam search is a non-optimal heuristic method, to further improve data locality while avoiding sequentialization, we iteratively optimize the mapping by applying local changes guided by the \textit{run time estimation model} to reconcile the two conflicting goals (Algorithm \ref{alg:estimation_model}). 
 The model takes  vertex pair s $\psi$ and the whole graph mapping $M=\{v:p| v \in V, p \in P\}$ as input. Each time, a vertex pair is selected to estimate the time to pass through the one-hop neighborhoods of these two vertices, called the \textit{partial run time}. Especially in case of \textit{sequentialization}, if vertices $(v_0, \dots, v_n)$ in a PE $p$ have incoming edges from the same vertex $u$, we denote $(v_0, \dots, v_n)$ as the \textit{collision set} and edges $\{(u,v)|v\in (v_0, \dots, v_n)\}$ as \textit{congested edges}.
 
\begin{figure}[t]
\centering
\includegraphics[width=0.6\textwidth]{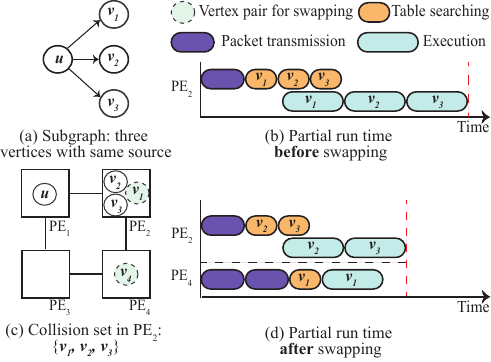}
\caption{Illustrative example of \textit{partial run time} estimation for PE with \textit{collision}.}
\label{figure:swap}
\end{figure}

    In each iteration, a vertex pair $(u,v)$ is evaluated for swapping (line 1). A vertex pair consists of a vertex in the current PE and another in a neighboring PE. To evaluate the benefit of swapping, we estimate partial run time before and after swapping. The partial run time is estimated by summing up the estimated run time of each connected edge of these two vertices (line 2). For one edge, the run time $t^e$ is estimated by three parts: the intra-PE addressing time (packet transmission) $t_{trans}$, vertex program execution time $t_{exe}$, and the intra-PE addressing time (table searching) $t_{tab}$. The packet transmission time is calculated by multiplying the time per hop and the number of hops (line 3). Especially when these two vertices are assigned to the same PE cluster but different slices, which means they cannot be loaded on-chip at the same time, an extra overhead $\epsilon$ is added (line 4). When the edge is one of the \textit{congested edges}, vertices need to be sequentially executed, as shown in  \fig \ref{figure:swap}. In this case,  we estimate the worst-case time with the assumption that this vertex is the last one in sequential processing (line 6). Otherwise, the partial run time is estimated by adding up the packet transmission time, table searching time and vertex program execution time (line 8). The same process is repeated to get the partial run time assuming a swapped location of the vertex pair (line 10).
    Finally, the vertex pair with the highest benefit is swapped if the partial run time decreases the most after swapping (line 13).

\begin{algorithm}[t]
\caption{\textsc{Flip} Estimation Model}
\label{alg:estimation_model}
\begin{algorithmic}[1]
\REQUIRE Mapping $M$=\{$v$:$p$\}, vertices pairs $\psi$, est. time for 1 hop trans. $t_{h}$ 
\ENSURE lowest-cost vertices pair ($u'$,$v'$) and its cost $c'$ for swapping
\FOR{($u$,$v$) in $\psi$}
    \FOR{$e$ in $u$.edges $\cup$ $v$.edges}
        \STATE $t_{trans}$ = $e$.length\_of\_route $\times$ $t_h$;  \COMMENT{Transmission time} 
        \STATE $t_{trans}$ += isSameClusterDiffChip($e$.srcPE, $e$.destPE)? $\epsilon$:0; 
        \IF{$e$ in \textit{congested edges}}
            \STATE $t^e$ = maximumTime(\textit{congested edges})
        \ELSE
            \STATE $t^e$ = $t_{trans}$ + $t_{tab}^e$+$t_{exe}$;\COMMENT{Add Table search and execution time}
        \ENDIF
        \STATE repeat to get ${t'}^e$, supposing swapped location of $u$ and $v$.
        \STATE $c$ = $t^e$ -${t'}^e$
    \ENDFOR
    \STATE update ($u'$,$v'$) and $c'$ if $c>c'$
\ENDFOR
\end{algorithmic}
\end{algorithm}

\begin{figure}[t]
\centering
\includegraphics[width=0.6\textwidth]{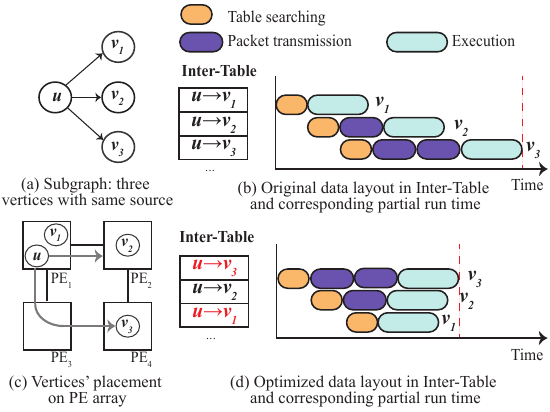}
\caption{Illustrative example of farthest-one-first strategy.}
\label{figure:Farthest_First}
\end{figure}
\subsection{Data Layout}
Besides mapping, an optimized data layout in Inter-Table can also reduce run time by changing the data access pattern.
The philosophy for efficient data layout is \textbf{"The farthest, the first"}.
For one vertex $u$ with three neighbors placed in different PEs, as in \fig \ref{figure:Farthest_First} (a, c). After $u$ updates its value, multiple packets heading to different destinations are issued sequentially with the order of entries in Inter-Table.
Without sorting the Inter-Table, the worst case may happen when the packet heading to the farthest destination is the last one generated (see \fig \ref{figure:Farthest_First} (b)). Meanwhile, with sorting, these three vertices can finish processing with less time (see \fig \ref{figure:Farthest_First} (d)).
An inefficient data layout will increase the overall processing time for all neighbors of a vertex and lead to under-utilization of the PE array.
Even though asynchronous execution does not have the straggler problem \cite{low2014graphlab, cipar2013solving} in bulk synchronous parallel execution, the routing to the farthest destination is highly likely to be a critical path in the whole execution and optimization is necessary.

\subsection{Support for Data Swapping}
\textsc{Flip} compiler is extended to support data swap by replicating the PE array into \texttt{$\lceil\frac{|V|}{P.capacity} \rceil$} copies. As data swapping occurs in a basic unit of 2$\times$2 PE cluster, only edges where the end vertices are mapped to the same PE cluster but different PE array copies are estimated to have higher transmission time. 

\begin{table}[t]
  \centering
  \caption{Comparison of \textsc{Flip} with CGRA-based accelerators.}
  \begin{tabular}{c c c c c a}
  \hline
        	& \textbf{PolyGraph}  & \textbf{Fifer}    & \textbf{HyCUBE\textsuperscript{*}}	& \textbf{RipTide}&  \textbf{\textsc{Flip}} \\
    		& \cite{dadu2021polygraph} 	  	& \cite{nguyen2021fifer} 		&  \cite{wang2019hycube}  & \cite{riptide} 	& (\textit{this work})\\
  \hline
 Goal 	& High Perf. & High Perf. & Low Pwr. & Ultra Low Pwr. & Low Pwr. \\
    
    \multirow{2}{*}{Target} & \multirow{2}{*}{Graph App.}  &  \multirow{2}{*}{Irreg. App.} & \multirow{2}{*}{Reg. Loop}  & \multirow{2}{*}{General App.} & Graph App. \& \\
    & & &  & &  Reg. Loop \\
    \hline\hline

    On-Chip Memory 	&512MB 	& 4.5MB	& 4KB & 256KB	& 32KB  \\
    Frequency 	& 1GHz 	& 2GHz	& 488MHz &  50MHz & 100MHz\\
    Technology (nm) 	& 28 & 22 	& 40 & sub-28 & 22\\

    \hline \hline
    Power (mW) 	&  2292 	& N/A	& 140	&  0.5\textsuperscript{+}	& 26  \\ 
    Area (mm$^2$) 	& 73	& 21	 & 3	& 0.3\textsuperscript{+} & 0.4 \\ 
\hline
     \multicolumn{6}{c}{\textsuperscript{*}Silicon implementation. \textsuperscript{+}PE array only, w/o on-chip memory.}
  \end{tabular}
  \label{table:sota_CGRA_details}
\end{table}

\section{Experimental Evaluation}
We evaluated \textsc{Flip} against a micro-controller unit (MCU) and a classic CGRA \cite{karunaratne2017hycube}, achieving up to 403$\times$ and 36$\times$ speedup respectively. We also compare \textsc{Flip} with PolyGraph \cite{dadu2021polygraph}, a state-of-the-art datacenter-scale dedicated graph processor based on CGRA, and demonstrate better performance per 
area. 
Table \ref{table:sota_CGRA_details} compares \textsc{Flip} with recent CGRA-based accelerators using reported numbers from the papers. PolyGraph~\cite{dadu2021polygraph} and Fifer~\cite{nguyen2021fifer} are state-of-the-art datacenter-scale CGRA-based accelerators. PolyGraph is dedicated to graph processing while Fifer supports irregular applications. But they have huge power (2.3W) and area footprints compared to \textsc{Flip} (26mW). Compared to the edge accelerators HyCUBE and RipTide, \textsc{Flip} has similar area/power but is more versatile supporting graph processing in addition to regular kernels. Note that RipTide area and power are for the PE array only and exclude 256KB on-chip memory, whereas other accelerators include on-chip memory in the estimation. 

\subsection{Evaluation Methodology}

\textbf{Baseline Architectures.}
For detailed quantitative comparison using the same workload, we evaluate \textsc{Flip} against two edge architectures: \underline{MCU}: a mature ARM Cortex-M4F core running at 64MHz,
and \underline{CGRA}: classic static-scheduled CGRA similar to \cite{wang2019hycube}.
Both classic CGRA and \textsc{Flip} have 8$\times$8 PE arrays running at 100MHz and 16KB on-chip SPM. Note that \textsc{Flip} has an additional
distributed on-chip storage (DRF and Inter/Intra-Tables). 


\noindent\textbf{Implementation.}
Both classic CGRA and \textsc{Flip} are implemented in System Verilog RTL~\cite{8299595} and synthesized on 22nm process using \textit{the Synopsys} toolchain to evaluate power and area. We implement \textsc{Flip} compiler for graph data mapping. Morpher compile \cite{wijerathne2022morpher} is used for DFG mapping on the classic CGRA. We build an in-house cycle-accurate simulator for the whole system to faithfully execute the applications on \textsc{Flip} architecture for faster evaluation of performance. Functional validation is performed through the simulator.  

 \begin{table}[t]
\caption{Workloads for Evaluation.}
  \label{table:workload}
  \centering
\begin{tabular}{ |c|c|c|  }
 \hline
 \textbf{Workload}  & \textbf{Description}  & \textbf{Time Complexity}\\\hline 
 \textbf{BFS}       & BFS level             & $O(|E|+|V|) $\\ \hline
 \textbf{SSSP}      & shortest path         & $O(|E|+|V|log|V|)$, $O(|V|^2)$\\ \hline
 \textbf{WCC}       & connected component   & $O(|E|+|V|) $\\ \hline 
\end{tabular}
\end{table}

\noindent\textbf{Workload.}
We choose the widely-used graph algorithms: Breadth-First Search (BFS), Single-Source Shortest Paths (SSSP), and Weakly Connected Components (WCC) for evaluation, as shown in Table \ref{table:workload}.
For SSSP, the solution with optimal time complexity $O(|E|+|V|log|V|)$ needs to use an advanced data structure, the priority queue built on binary search tree. Classic CGRAs designed for regular computation-intensive kernels cannot support such dynamic and complex data structure.
Therefore, we implemented the algorithm with $O(|V|^2)$ time complexity for SSSP on classic CGRA, while for MCU, the optimal one is employed. For classic CGRA, to iterate over one vertex, 34/38 operations are needed in BFS and WCC. In SSSP, two kernels with 10/31 operations will be mapped for vertex searching and updating.
For \textsc{Flip}, the number of instructions for processing one vertex is 4/5/5 for WCC, BFS and SSSP when the vertex's properties are updated. If there is no update, only 2/4/4 instructions are executed. In other words, the computation cost is minimized in \textsc{Flip} as we do not need to generate memory addresses for load/store instructions. 

\noindent\textbf{Datasets.}
Table \ref{table:dataset} shows the graphs we use to evaluate \textsc{Flip}. The original road network of California and San Francisco are retrieved from the Standard Large Network Dataset Collection (SNAP) \cite{leskovec2014snap}.
As edge devices rarely use the entire road network of a whole state (city), we construct our datasets with small subgraphs obtained by BFS on the original road networks with random seeds.
Additionally, a data set of synthetic graphs with short diameters is built by randomly connecting two vertices.
To investigate the performance of \textsc{Flip} for different kinds of graphs, we constructed five types of graphs: \textit{small road networks} (SRN), \textit{large road network}s (LRN), \textit{tree}s (Tree), \textit{synthetic graphs} (Syn) and extra large road networks (Ext. LRN). The first four types of graphs can be stored fully on-chip and are used for performance evaluation. The last data set is only used for scalability evaluation, where runtime data swapping from off-chip memory is required.

\noindent\textbf{Performance Evaluation Methodology.}
To evaluate the performance of \textsc{Flip}, we run all the applications against the graph datasets in our simulator. We report the number of cycles when the application terminates for each run. For all graphs other than Tree, we use 100 random vertices as the source vertex in each run and report the average. For Tree, applications only start with the root vertex.

{
\begin{table}[t]
\caption{Graph characteristics for varying CGRA sizes.}
  \label{table:dataset}
  \centering
\begin{tabular}{  c c c c c c   }
 \hline 
\textbf{Group} & \textbf{Type} & \textbf{Diameter} &\textbf{\# Graphs}  & $\mathbf{\lvert V\rvert}$ & $\mathbf{\lvert E\rvert}$ \\ \hline
\textbf{Tree} & Directed & High & 100 & 256 & 255 \\ 
\textbf{SRN} & Undirected & High & 100&[64,107]  &[146,278] \\  
\textbf{LRN} & Undirected & High & 100& 256  & [584,898] \\  
\textbf{Syn.} & Directed & Low & 100 & 256 & 768 \\ 
\textbf{Ext. LRN} &  Undirected & High & 10 & 16k & [44k,50k] \\ \hline 
\end{tabular}
\end{table}
}

\subsection{Experimental Results}
\subsubsection{\bf Performance}
\begin{figure}[t]
\centering
    \includegraphics[width=5in]{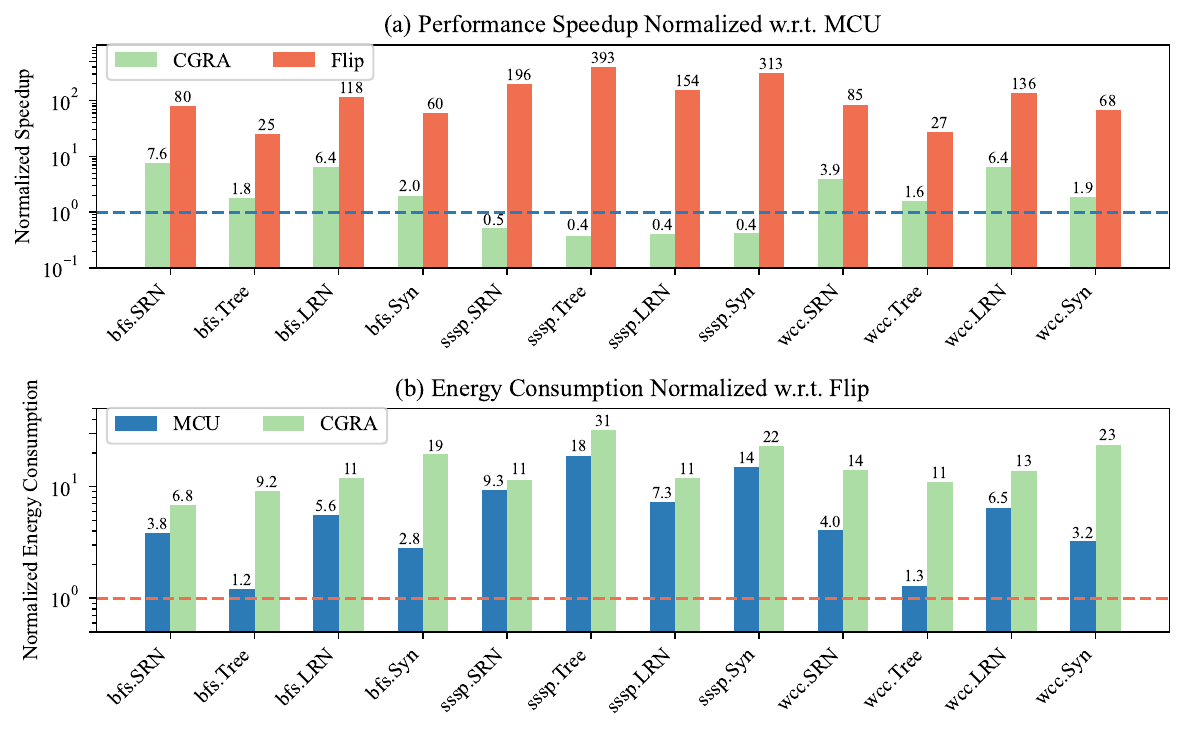}
	\caption{Performance, energy comparison in \textit{log} scale. Energy consumption for MCU is for the core only and excludes on-chip memory.}
	\label{figure:perf_energy}
\end{figure}

\fig \ref{figure:perf_energy}(a) shows the performance normalized to MCU for three different graph applications on four data sets.
 Note the logarithmic scale for the Y-axis.
\textsc{Flip} outperforms the baseline architectures for all workloads on all types of graphs. It achieves a 25$\times$-393$\times$ speedup compared to MCU. Eliminating the impact of 
non-optimal implementation, \textsc{Flip} achieves an 11$\times$-36$\times$ speedup compared to classic CGRA on BFS and WCC. 
Classic CGRA cannot expose sufficient parallelism within each loop iteration, given the small code snippet for processing each vertex and complex dependencies among vertices.
In contrast, \textsc{Flip} improves the performance by exploiting data parallelism at runtime and enabling multiple active vertices to execute in parallel. 

Classic CGRA outperforms MCU in most cases, because MCU is a 5-stage single-issue in-order core, while these CGRA can compute and access memory in parallel with multiple PEs. MCU performs better than CGRA on SSSP due to the non-optimal algorithm implemented on classic CGRA.

\begin{table}[t]
\caption{Performance-power-area comparison.}
\label{table:power_area_efficiency}
\centering
\begin{tabular}{l r r r r r r}
 \hline 
       	& \textbf{MTEPS} & \textbf{Power}	& \textbf{Area} &  \textbf{Power Efficiency} & \textbf{Area Efficiency} &\textbf{Technology} \\
       	&  & (mW) & (mm$^2$) & MTEPS/mW & MTEPS/mm$^2$ & nm   \\\hline
MCU (LRN) & 1.1 & 0.78\textsuperscript{$\ddagger$} & 0.03\textsuperscript{$\ddagger$} & 1.41 & 39 & 22\\
CGRA (LRN)	& 7.1  & 17	& 0.32	& 0.43	& 22 & 22\\
\textsc{Flip} (LRN)    & 158   & \cellcolor{yellow} 26 	& \cellcolor{yellow} 0.37 &\textbf{6.12} & \textbf{424} & 22\\ \hline
PolyGraph (from~\cite{dadu2021polygraph})   & 13,845  & \cellcolor{yellow} 2292   & \cellcolor{yellow}72.56	& 6.04	& 191 & 28 \\
 \hline
\multicolumn{7}{l}{Fifer~\cite{nguyen2021fifer} \& RipTide~\cite{riptide} \ \ MTEPS is not available from papers}\\\hline
\multicolumn{6}{c}{ \textsuperscript{$\ddagger$}On-chip memory excluded.}\\
\end{tabular}
\end{table}
\subsubsection{\bf Power, Area and Energy}
 As shown in Table \ref{table:power_area_efficiency}, with the same size of PE array, \textsc{Flip} has only 19\% more area but 53\% more power than classic CGRA due to extra logic on the NoC and buffer queues at the ports to support dynamic routing. 
Despite higher power, \fig \ref{figure:perf_energy}(b) shows that \textsc{Flip} requires only 5\%-82\% and 3\%-15\% energy compared to the MCU and classic CGRA, respectively, 
due to dramatically improved performance. Note that this comparison is biased towards MCU, as MCU energy is for the core only and does not include on-chip memory, whereas \textsc{Flip} energy includes a total 32KB on-chip memory (16KB SPM and 16KB distributed memory inside PE array).  Table \ref{table:breakdown} shows the power and area breakdown of \textsc{Flip}. Memory components consume 92.76\% of the power and 88.19\% of the area. Moreover, \textsc{Flip} keeps a 32-entries instruction memory, taking 20\% the power and area, to accommodate most kernels in operation-centric mode. 

\subsubsection{\bf Efficiency}
Table \ref{table:power_area_efficiency} also shows power and area efficiency of \textsc{Flip} compared with classic CGRA and a state-of-the-art CGRA-based large-scale graph accelerator, PolyGraph \cite{dadu2021polygraph}. We use the absolute values mentioned in the original manuscript without scaling to a common setup. Because of the inconsistency in setup details, it is not feasible to scale to a common setup condition. Hence we use this table to roughly place \textsc{Flip} among the current state of the art.
The standard graph processing metric is used: Million~Traversed~Edges~Per Second~(MTEPS).
The performance is evaluated on WCC workload for all architectures. We report the average performance of PolyGraph running WCC on two unordered road networks, \texttt{rdUSE} and \texttt{rdUSW},  retrieved directly from~\cite{dadu2021polygraph}. 
Compared to PolyGraph, \textsc{Flip} achieves similar power efficiency and 2.2$\times$ area efficiency. This is because \textsc{Flip} compiler captures better data locality reducing communication costs. Designed for edge-scale problems, we judiciously avoid complex components (e.g., scheduler and task coalescer) in \textsc{Flip}.  PolyGraph uses complex controllers for data-center scale problems that are not feasible under stringent power/area constraints at the edge. \textsc{Flip} uses only 1.1\% and 0.5\% of the area and power of PolyGraph.
\begin{table}[t]
\caption{Power and Area Breakdown for \textsc{Flip}.}
  \label{table:breakdown}
  \centering
\begin{tabular}{ |c|c|c|c|  }
    \hline
\textbf{Type} &  \textbf{Component} & \textbf{Power (mW)}& \textbf{Area (mm$^2$)}\\
\hline
    \textbf{Interconnect}            & \textbf{Switch Allocator}         & 0.08 (0.31\%) &  0.006 (1.60\%) \\ 
    \hline
     \textbf{Compute Unit }     & \textbf{ALU}           &0.01 (0.04\%)	& 0.004 (0.97\%) \\
    \hline
    \multirow{8}{*}{\textbf{Memory}}   & \textbf{Inter-Table}    & 5.91(22.90\%)  &  0.073 (19.56\%) \\
    \cline{2-4}
                                        & \textbf{Intra-Table}    &5.39 (20.89\%)	& 0.065 (17.34\%) \\
    \cline{2-4}
                                        & \textbf{ALUout Buffer}  & 0.07 (0.27\%) &  0.021 (5.60\%)\\
    \cline{2-4}
                                        & \textbf{ALUin Buffer}   & 1.05 (4.07\%) &  0.011 (2.89\%) \\
     \cline{2-4}
                                        & \textbf{Memory Buffer}  & 0.75 (2.90\%) &  0.008 (2.90\%) \\
     \cline{2-4}
                                & \textbf{Input Buffer}         & 4.02 (15.57\%) & 0.055 (14.74\%)\\
    \cline{2-4}
                                & \textbf{DRF}            & 1.75 (6.77\%) &  0.021 (5.72\%) \\   
     \cline{2-4}
                                & \textbf{Instruction Memory}  & 4.89 (18.96\%)  &  0.074 (19.74\%) \\ 
    \cline{2-4}
                                & \textbf{Slice ID Register}  & 0.11 (0.42\%) & 0.001 (0.36\%) \\ 
    \hline
    \multicolumn{2}{|c|}{\textbf{Additional Logic}} & 1.78 (6.89\%) & 0.034 (9.24\%)\\
    \hline \hline
    \multicolumn{2}{|c|}{\textbf{Total}}  & 25.79 & 0.373\\
    \hline
\end{tabular}
\end{table}
\begin{figure}[t]
\centering
\includegraphics[width=3.6in, trim={0 0 0 0},clip]{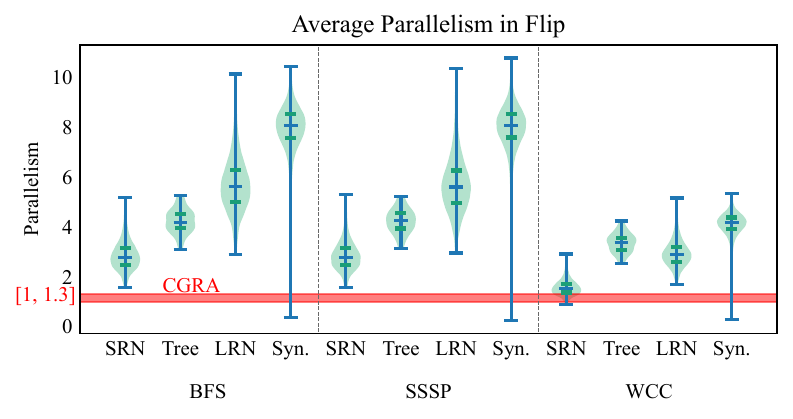}
\caption{Average parallelism for all runs on operation-centric CGRA and \textsc{Flip}.}
\label{figure:parallelism}
\end{figure}
\subsubsection{\bf Parallelism} \label{section:parallelism}
\fig \ref{figure:parallelism} shows the comparison of parallelism among classic CGRA and \textsc{Flip}.
Parallelism is defined as the number of active vertices at the same time. 
Operation-centric CGRAs were limited to processing fixed-size, tiny batches of vertices due to uncertain data dependencies, but they can exploit instruction parallelism within the computation for a vertex. As shown in the red area of Figure \ref{figure:parallelism}, the actual average parallelism is only 1 to 1.3 for different unroll degrees (1-4).

In contrast, our approach accommodates variable-sized data batches, potentially handling considerably larger volumes.

As shown in \fig \ref{figure:parallelism}, \textsc{Flip} achieves an average parallelism of 5.0-7.6 in most runs (25\% quantile) when storage is fully utilized (LRN and Syn). Note that graph processing has naturally low parallelism at the beginning and end of execution. With a starting vertex in the center of the graph, the average parallelism can reach 10.4.
However, the parallelism is fixed at a low level for classic CGRA.
WCC performs worse than the other two applications on \textsc{Flip} due to the relatively minor computation demand, making the performance bounded by communication.
For synthetic graphs, as not every vertex can reach all other vertices, the average parallelism in some rare runs could be low.

\subsubsection{\bf Scalability}
\label{section:scalability}
\begin{figure}[t]
\centering
\includegraphics[width=3.4in]{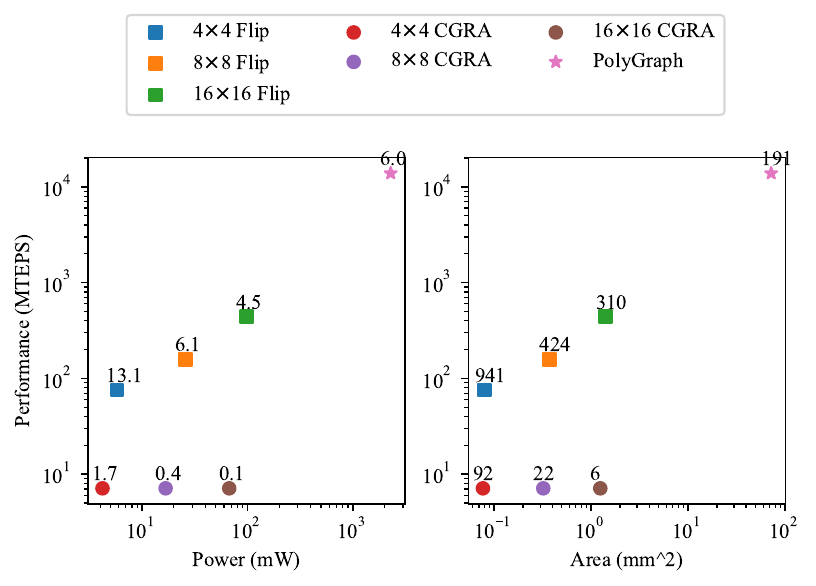}
\caption[PPA]{Performance-power/-area comparison. Performance is evaluated for WCC with road networks that fully utilize the on-chip distributed memory. Memory size inside one PE remains constant during scaling. The number beside the node denotes the power/area efficiency.}
\label{figure:power_area_efficiency}
\end{figure}

\textsc{Flip} is capable of bringing in data at runtime from a 256KB off-chip main memory for larger problem sizes. Evaluated on Ext. LRN, the performance drops due to the overhead of data swapping at runtime but still greatly outperforms classic CGRA (5.7$\times$ throughput) and MCU (49.1$\times$ throughput). With a homogeneous PE array, \textsc{Flip} can be easily scaled up. However, as shown in \fig \ref{figure:power_area_efficiency}, by linearly increasing PE array size and memory size, \textsc{Flip} shows some performance drop. This is because, for road networks, a larger graph denotes a longer diameter, requiring more hops to reach all vertices in the graph. Therefore, although the dataset and PE array are scaled at the same rate, the total execution time still grows, causing performance degradation. 

\subsubsection{\bf Compilation Time and Quality}
\label{section:compilation_n_quality}
\begin{figure}[t]
	\centering
\includegraphics[width=4in]{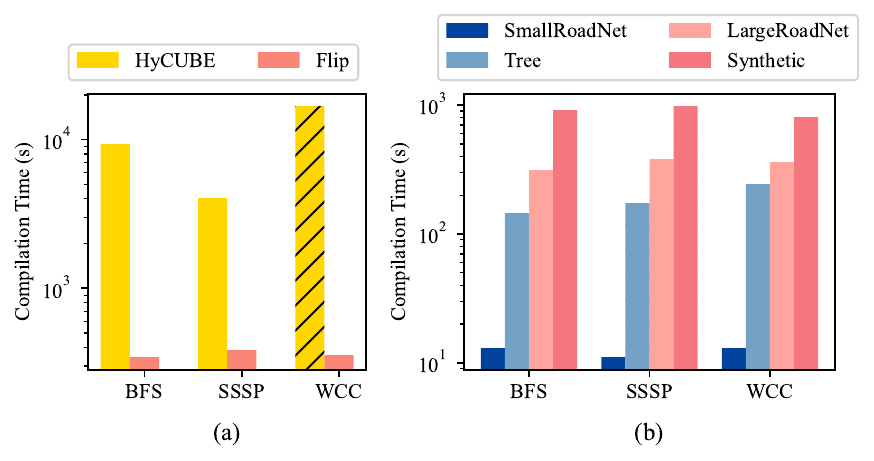}	
 \caption{ (a) Compilation time for three CGRAs. Average compilation time for all graphs is used for \textsc{Flip}.
    (b) Compilation time of \textsc{Flip} with different graph sizes.}
	\label{figure:compilation_time}
\end{figure}

The compiler for classic CGRA generates mapping along the spatial and temporal dimensions, and hence has long compilation time. In contrast, the \textsc{Flip} compiler only needs to map the graph vertices on PE array without any consideration of the temporal dimension. Due to the limited exploration space, the \textsc{Flip} compiler generates a valid mapping in less than 1\% to 10\% compilation time of conventional CGRAs, as shown in \fig \ref{figure:compilation_time}. The Y-axis in this graph has logarithmic scale. However, compared with graph partition methods applied on data-center scale problems\cite{zhang2018graphp, dadu2021polygraph, ahn2015scalable} where a maximum $O(|V|+|E|)$ complexity is allowed, \textsc{Flip} compiler provides a finer placement considering the routing distance as well as parallelism at the cost of time complexity. Here we analyze the complexity of \textsc{Flip} compiler's two phases. The first phase is to use beam search with routing length as the cost to find an initial placement. For one node, the routing length is calculated by Manhattan distance for all connected edges with $O(\frac{|E|}{|V|})$ complexity, and the beam search has time complexity $O(k|V|)$ where $k$ is set to 10 in \textsc{Flip} compiler. So, there is a total $O(|E|)$ complexity in first phase. The second phase is to optimize the placement locally until stabilization. Therefore, the time complexity for one iteration is $O(\frac{|V|+C|E|}{|P|})$, where $|P|$ is the number of PEs, $C$ is the number of vertices a PE can accommodate. Table \ref{table:compiler_complexity} shows the breakdown of the time complexity.

\begin{table}[t]
\centering

\caption{Time Complexity Breakdown for \textsc{Flip} Compiler.}
\label{table:compiler_complexity}
\begin{tabular}{ l l l l c }
\hline 
\multicolumn{4}{l}{\textbf{Process}} & \textbf{Time Complexity} \\
\hline\hline
\multicolumn{4}{l}{\textbf{Initial Mapping}}  &  $O(k|V|)$ \\
\multicolumn{4}{l}{\textbf{Local Optimization (One Iteration)}} & $O(\frac{|V|+C|E|}{|P|})$ \\
&\multicolumn{3}{l}{get neighboring PEs of a random PE} & $O(\frac{|V|}{|P|C})$ \\
& & \multicolumn{2}{l}{get collision set} & $O(C)$\\
& & \multicolumn{2}{l}{get candidate vertices pairs} & $O(C^2)$\\
& & & time estimation for all edges of vertices pair & $O(\frac{|E|}{|V|})$\\
\hline 
\multicolumn{5}{c}{ {\footnotesize $k$: beam width in beam search. $C$: capacity of one PE. $|P|$: number of PEs.}}
\end{tabular}

\end{table}


To evaluate the quality of mapping, we report statistics that indirectly capture the success of compiler in minimizing routing length and contentions for resources in Table \ref{table:routing_length}. 
The \textsc{Flip} compiler generates mappings with an average routing length of less than 1 for road networks. Recall that we might map neighboring vertices to the same PE (i.e., routing length 0) if it does not hurt parallelism. 

The table also reports the contention for ALU when multiple active vertices are mapped to the same PE (ALUin Buffer Depth) and contention for routing resources (Pkt. Wait Time). The contention for ALU is negligible, while the waiting time for NoC resources is less than 10 cycles even for large networks, demonstrating the efficacy of the compiler.

\begin{table}[t]
\centering
\caption{SSSP for different graph sizes on \textsc{Flip}}
\label{table:routing_length}
\begin{tabular}{ |c|c|c|c|c|  }
 \hline
     \textbf{Graph Group}  & \textbf{SRN}& \textbf{LRN}& \textbf{Tree} & \textbf{Syn.}\\
 \hline
    \textbf{Avg. Routing Length} &  0.63 & 0.76 & 0.55  & 2.46 \\
 \hline
     \textbf{Pkt. Wait Time (cycle)} & 7.8 &  9.6& 5.1 & 7.9\\ \hline
  \textbf{ALUin Buffer Depth}  & 0.04 & 0.08 & 0.03 & 0.14 \\
 \hline 
\end{tabular}
\end{table}

\section{Related Work}
We discuss related work from three perspectives: conventional operation-centric CGRAs, dedicated graph accelerators, and processing-in-memory acceleration. Three highly-correlated works are compared in detail.
Note that the performance of a graph accelerator also depends on the parallelism available in the input graph data. The real-world road networks we use in \textsc{Flip}, which serves as an edge device, have much less parallelism compared to the synthetic graphs. Therefore, the power efficiency (MTEPS/Watt) should not be directly compared.
\subsection{Conventional Operation-centric CGRAs}
Conventional CGRAs exploit the Instruction Level Parallelism~(ILP) to accelerate loop kernels. 
Many recently proposed CGRAs \cite{wang2019hycube, nowatzki2017stream, wijerathne2019cascade, yuan2021dynamic, nguyen2021fifer, dadu2022systematically} optimize performance from various perspectives.  OpenCGRA \cite{tan2020opencgra, tan2021opencgra} provides an open-source full-stack unified framework to allow rapid design space exploration of CGRA. 
 \cite{liu2022overgen, weng2020dsagen, tan2021aurora, tan2022asap, bandara2022revamp} optimize the design of CGRA architecture for a specific set of applications through automatic design space exploration.
Moreover, to handle common conditional branches in graph applications, CGRAs \cite{hamzeh2014branch, karunaratne20194d}  require predication support by simultaneously mapping both paths of each conditional branch, necessitating additional hardware resources to store both paths.
Despite all the efforts, all works still adhere to the conventional operation-centric approach. However, these methods still follow routine data-flow mapping with data streaming in exploiting ILP as well as regular independent data-level parallelism, which is not suitable for graph applications, while \textsc{Flip} directly maps the data and exploits the irregular data-level parallelism in the presence of rich dependencies.
Furthermore, conventional CGRAs need to calculate data addresses for each access after loading graph data into on-chip memory. 
Although some CGRAs\cite{wijerathne2019cascade, nowatzki2017stream,weng2020hybrid, dadu2021polygraph, riptide} introduce address generation units, the operation-centric design still leads to poor performance for graph applications with irregular memory access.

\noindent \textbf{CGRAs with Dynamic Routing.}
Several CGRAs \cite{swanson2003wavescalar, mishra2006tartan, waingold1997baring} are equipped with dynamic routing to support irregular and unpredictable communication. However, dynamic routing is only available at a top level among clusters of processing elements (at least 128 PEs in one cluster) \cite{swanson2003wavescalar, mishra2006tartan} or RISC-like processors (32KB memory in one tile) \cite{waingold1997baring}. This is the same level as the whole \textsc{Flip} chip. Inside a PE cluster, \cite{swanson2003wavescalar, mishra2006tartan} still use static routing scheduled by the compiler. Therefore, when scaled down to the same range as \textsc{Flip}, these architectures can hardly benefit from dynamic routing.

\subsection{Dedicated Graph Accelerators}
A typical design for ASIC-based graph processors \cite{ham2016graphicionado, ozdal2016energy, zhou2017tunao, ayupov2017template} is to build hardware modules separately for the three phases in the \textit{Gather-Apply-Scatter} model.
A recent work by Luke et al. \cite{everson2021time} proposes simulating the pathfinder on a mesh network with signal passing through NoC, but it is limited to mesh-like road networks. 
FPGA-based graph processors \cite{kapre2015custom, zhou2018fpga, lei2015fpga} achieve high throughput by exploiting the Multiple Instruction Single Data execution pattern and large accumulated bandwidth of on-chip memory. 
Below we introduce the difference between \textsc{Flip} and two highly-correlated works: PolyGraph\cite{dadu2021polygraph} and GraphPulse\cite{rahman2020graphpulse}.
\\\textbf{Comparison between PolyGraph and \textsc{Flip}.}
PolyGraph\cite{dadu2021polygraph, dadu2022systematically} is a state-of-the-art CGRA-based graph processor that allows modular integration of specialization features (e.g., synchronization frequency, scheduling strategy, etc.) for each graph processing variant (e.g., graph diameter, graph density, graph algorithm). It comprises a 4$\times$4 core array connected by a mesh NoC. Each core has a control core for configuration and scheduling, a 5$\times$4 CGRA array for computation, and a task management system able for task caching, coalescing, and dispatching. It also features a comprehensive data access system for accurately retrieving data from local or other PEs without conflicts and offloading address computation to CGRA.
PolyGraph can also be easily extended for general-purpose applications as it embeds classic operation-centric CGRA in each core. 
However, PolyGraph's design cannot be directly deployed at the edge due to its complex memory access and task scheduling systems.
Even one core of PolyGraph is 5.5$\times$ the power and 11.9$\times$ area of \textsc{Flip}.
Additionally, PolyGraph uses CGRA in a classic way, while our approach innovatively alters the architecture to enable dual-mode execution. 
Finally, the execution inside each core in PolyGraph is managed by the task scheduler and control core. In other words, it is program-controlled. In contrast, \textsc{Flip} uses the data itself to trigger the subsequent execution in each PE, which is data-driven.
\\\textbf{Comparison between GraphPulse and \textsc{Flip}.}
GraphPulse\cite{rahman2020graphpulse} is an ASIC-based graph processor for efficient asynchronous graph processing. 
It bypasses the overhead of synchronization and frontier tracking in the Bulk Synchronous Parallel (BSP) processing model by employing an event-driven asynchronous processing model.
In GraphPulse, each event encompasses the destination vertex ID and the changed vertex attribute, similar to \textsc{Flip}. 
However, \textsc{Flip} allows data to be sent directly to the destination PE rather than being sent to a centralized event buffer and dispatched by a centralized scheduler. 
Although a centralized event coalescer helps reduce events heading to the same destination and subsequently decreases memory access, the coalescer is often area- and power-intensive, requiring a minimum size of the number of vertices in a graph multiplied by vertex attribute size for $O(1)$ access time. This makes GraphPulse unsuitable for edge devices with stringent power and area constraints. 

\subsection{Processing in Memory}
Processing-in-Memory (PIM) is an effective technique that mitigates data movement costs by integrating processing units within memory.
Graph processors enabled by PIM primarily utilize two technologies: ReRAM\cite{wong2012metal} and Hybrid Memory Cube (HMC)\cite{pawlowski2011hybrid},  which offer energy-efficient high-bandwidth memory access compared to conventional DRAM.
GRAPHR \cite{song2018graphr}, the first ReRAM-based graph processing accelerator, proposes to process the graphs in the SpMV format with crossbars.
Tesseract \cite{ahn2015scalable} is the pioneer in applying HMC technology for graph processing.
Solutions such as GraphP\cite{zhang2018graphp}, GraphQ, \cite{10.1145/3352460.3358256} and GraphH \cite{8328836} optimize inter-cube communication through graph partition, reordering processing order, and efficient NoC. 
\\\textbf{Comparison between Tesseract and \textsc{Flip}.}
Tesseract\cite{ahn2015scalable} constitutes a 4$\times$4 HMC array with 4$\times$8 vaults in each HMC. Each vault contains an in-order core for computation and task management, a data prefetcher, a data buffer, and a task queue, cumulatively facilitating up to 512 cores and 128GB of storage. Due to the constraints of large problem sizes, Tesseract employs an index-based graph partitioning method with low time complexity, which however overlooks the graph structure. In contrast, the \textsc{Flip} compiler, optimized for smaller problem sizes coupled with limited memory in each PE,   takes the graph structure into consideration for graph partitioning and mapping. Moreover, \textsc{Flip} compiler addresses the sequentialization problem which is typically not encountered in PIM-enabled graph processors, as each vault can store a significant portion of the active frontier, thereby minimizing the likelihood of staleness.
\section{Conclusions} 
CGRAs are prominent accelerators but perform poorly on applications with rich control and data dependencies, especially graph processing, due to static scheduling. \textsc{Flip} is a novel CGRA architecture that is good at graph processing at the edge while still supporting compute-intensive regular kernels. It adopts a novel data-centric model where the graph data (vertices) are mapped onto the PEs and the data dependencies (edges) are routed. \textsc{Flip} achieves up to 393$\times$ speedup compared to micro-controller and 36$\times$ speedup compared to conventional CGRA accelerator on graph applications. Compared to large-scale graph processor\cite{dadu2021polygraph}, \textsc{Flip} has similar energy efficiency and 2.2x better area efficiency at about 1\% of the power/area cost.

\bibliographystyle{ACM-Reference-Format}
\bibliography{main}


\begin{thebibliography}{62}


\ifx \showCODEN    \undefined \def \showCODEN     #1{\unskip}     \fi
\ifx \showDOI      \undefined \def \showDOI       #1{#1}\fi
\ifx \showISBNx    \undefined \def \showISBNx     #1{\unskip}     \fi
\ifx \showISBNxiii \undefined \def \showISBNxiii  #1{\unskip}     \fi
\ifx \showISSN     \undefined \def \showISSN      #1{\unskip}     \fi
\ifx \showLCCN     \undefined \def \showLCCN      #1{\unskip}     \fi
\ifx \shownote     \undefined \def \shownote      #1{#1}          \fi
\ifx \showarticletitle \undefined \def \showarticletitle #1{#1}   \fi
\ifx \showURL      \undefined \def \showURL       {\relax}        \fi
\providecommand\bibfield[2]{#2}
\providecommand\bibinfo[2]{#2}
\providecommand\natexlab[1]{#1}
\providecommand\showeprint[2][]{arXiv:#2}

\bibitem[829(2018)]%
        {8299595}
 \bibinfo{year}{2018}\natexlab{}.
\newblock \showarticletitle{IEEE Standard for SystemVerilog--Unified Hardware Design, Specification, and Verification Language}.
\newblock \bibinfo{journal}{\emph{IEEE Std 1800-2017 (Revision of IEEE Std 1800-2012)}} (\bibinfo{year}{2018}), \bibinfo{pages}{1--1315}.
\newblock
\urldef\tempurl%
\url{https://doi.org/10.1109/IEEESTD.2018.8299595}
\showDOI{\tempurl}


\bibitem[Ahn et~al\mbox{.}(2015)]%
        {ahn2015scalable}
\bibfield{author}{\bibinfo{person}{Junwhan Ahn}, \bibinfo{person}{Sungpack Hong}, \bibinfo{person}{Sungjoo Yoo}, \bibinfo{person}{Onur Mutlu}, {and} \bibinfo{person}{Kiyoung Choi}.} \bibinfo{year}{2015}\natexlab{}.
\newblock \showarticletitle{A scalable processing-in-memory accelerator for parallel graph processing}. In \bibinfo{booktitle}{\emph{Proceedings of the 42nd Annual International Symposium on Computer Architecture}}. \bibinfo{pages}{105--117}.
\newblock


\bibitem[Ayupov et~al\mbox{.}(2017)]%
        {ayupov2017template}
\bibfield{author}{\bibinfo{person}{Andrey Ayupov}, \bibinfo{person}{Serif Yesil}, \bibinfo{person}{Muhammet~Mustafa Ozdal}, \bibinfo{person}{Taemin Kim}, \bibinfo{person}{Steven Burns}, {and} \bibinfo{person}{Ozcan Ozturk}.} \bibinfo{year}{2017}\natexlab{}.
\newblock \showarticletitle{A template-based design methodology for graph-parallel hardware accelerators}.
\newblock \bibinfo{journal}{\emph{IEEE Transactions on Computer-Aided Design of Integrated Circuits and Systems}} \bibinfo{volume}{37}, \bibinfo{number}{2} (\bibinfo{year}{2017}), \bibinfo{pages}{420--430}.
\newblock


\bibitem[Bandara et~al\mbox{.}(2022)]%
        {bandara2022revamp}
\bibfield{author}{\bibinfo{person}{Thilini~Kaushalya Bandara}, \bibinfo{person}{Dhananjaya Wijerathne}, \bibinfo{person}{Tulika Mitra}, {and} \bibinfo{person}{Li-Shiuan Peh}.} \bibinfo{year}{2022}\natexlab{}.
\newblock \showarticletitle{REVAMP: a systematic framework for heterogeneous CGRA realization}. In \bibinfo{booktitle}{\emph{Proceedings of the 27th ACM International Conference on Architectural Support for Programming Languages and Operating Systems}}. \bibinfo{pages}{918--932}.
\newblock


\bibitem[Besta et~al\mbox{.}(2021)]%
        {10.1145/3466752.3480133}
\bibfield{author}{\bibinfo{person}{Maciej Besta}, \bibinfo{person}{Raghavendra Kanakagiri}, \bibinfo{person}{Grzegorz Kwasniewski}, \bibinfo{person}{Rachata Ausavarungnirun}, \bibinfo{person}{Jakub Ber\'{a}nek}, \bibinfo{person}{Konstantinos Kanellopoulos}, \bibinfo{person}{Kacper Janda}, \bibinfo{person}{Zur Vonarburg-Shmaria}, \bibinfo{person}{Lukas Gianinazzi}, \bibinfo{person}{Ioana Stefan}, \bibinfo{person}{Juan~G\'{o}mez Luna}, \bibinfo{person}{Jakub Golinowski}, \bibinfo{person}{Marcin Copik}, \bibinfo{person}{Lukas Kapp-Schwoerer}, \bibinfo{person}{Salvatore Di~Girolamo}, \bibinfo{person}{Nils Blach}, \bibinfo{person}{Marek Konieczny}, \bibinfo{person}{Onur Mutlu}, {and} \bibinfo{person}{Torsten Hoefler}.} \bibinfo{year}{2021}\natexlab{}.
\newblock \showarticletitle{SISA: Set-Centric Instruction Set Architecture for Graph Mining on Processing-in-Memory Systems}. In \bibinfo{booktitle}{\emph{MICRO-54: 54th Annual IEEE/ACM International Symposium on Microarchitecture}} (Virtual Event, Greece) \emph{(\bibinfo{series}{MICRO '21})}. \bibinfo{publisher}{Association for Computing Machinery}, \bibinfo{address}{New York, NY, USA}, \bibinfo{pages}{282–297}.
\newblock
\showISBNx{9781450385572}
\urldef\tempurl%
\url{https://doi.org/10.1145/3466752.3480133}
\showDOI{\tempurl}


\bibitem[Besta et~al\mbox{.}(2019)]%
        {besta2019graph}
\bibfield{author}{\bibinfo{person}{Maciej Besta}, \bibinfo{person}{Dimitri Stanojevic}, \bibinfo{person}{Johannes De~Fine Licht}, \bibinfo{person}{Tal Ben-Nun}, {and} \bibinfo{person}{Torsten Hoefler}.} \bibinfo{year}{2019}\natexlab{}.
\newblock \showarticletitle{Graph processing on fpgas: Taxonomy, survey, challenges}.
\newblock \bibinfo{journal}{\emph{arXiv preprint arXiv:1903.06697}} (\bibinfo{year}{2019}).
\newblock


\bibitem[Bisson et~al\mbox{.}(2016)]%
        {7234934}
\bibfield{author}{\bibinfo{person}{Mauro Bisson}, \bibinfo{person}{Massimo Bernaschi}, {and} \bibinfo{person}{Enrico Mastrostefano}.} \bibinfo{year}{2016}\natexlab{}.
\newblock \showarticletitle{Parallel Distributed Breadth First Search on the Kepler Architecture}.
\newblock \bibinfo{journal}{\emph{IEEE Transactions on Parallel and Distributed Systems}} \bibinfo{volume}{27}, \bibinfo{number}{7} (\bibinfo{year}{2016}), \bibinfo{pages}{2091--2102}.
\newblock
\urldef\tempurl%
\url{https://doi.org/10.1109/TPDS.2015.2475270}
\showDOI{\tempurl}


\bibitem[Buluc et~al\mbox{.}(2017)]%
        {buluc2017distributedmemory}
\bibfield{author}{\bibinfo{person}{Aydin Buluc}, \bibinfo{person}{Scott Beamer}, \bibinfo{person}{Kamesh Madduri}, \bibinfo{person}{Krste Asanovic}, {and} \bibinfo{person}{David Patterson}.} \bibinfo{year}{2017}\natexlab{}.
\newblock \bibinfo{title}{Distributed-Memory Breadth-First Search on Massive Graphs}.
\newblock
\newblock
\showeprint[arxiv]{1705.04590}~[cs.DC]


\bibitem[Cipar et~al\mbox{.}(2013)]%
        {cipar2013solving}
\bibfield{author}{\bibinfo{person}{James Cipar}, \bibinfo{person}{Qirong Ho}, \bibinfo{person}{Jin~Kyu Kim}, \bibinfo{person}{Seunghak Lee}, \bibinfo{person}{Gregory~R Ganger}, \bibinfo{person}{Garth Gibson}, \bibinfo{person}{Kimberly Keeton}, {and} \bibinfo{person}{Eric Xing}.} \bibinfo{year}{2013}\natexlab{}.
\newblock \showarticletitle{Solving the straggler problem with bounded staleness}. In \bibinfo{booktitle}{\emph{14th Workshop on Hot Topics in Operating Systems (HotOS XIV)}}.
\newblock


\bibitem[Dadu et~al\mbox{.}(2021)]%
        {dadu2021polygraph}
\bibfield{author}{\bibinfo{person}{Vidushi Dadu}, \bibinfo{person}{Sihao Liu}, {and} \bibinfo{person}{Tony Nowatzki}.} \bibinfo{year}{2021}\natexlab{}.
\newblock \showarticletitle{Polygraph: Exposing the value of flexibility for graph processing accelerators}. In \bibinfo{booktitle}{\emph{2021 ACM/IEEE 48th Annual International Symposium on Computer Architecture (ISCA)}}. IEEE, \bibinfo{pages}{595--608}.
\newblock


\bibitem[Dadu et~al\mbox{.}(2022)]%
        {dadu2022systematically}
\bibfield{author}{\bibinfo{person}{Vidushi Dadu}, \bibinfo{person}{Sihao Liu}, {and} \bibinfo{person}{Tony Nowatzki}.} \bibinfo{year}{2022}\natexlab{}.
\newblock \showarticletitle{Systematically understanding graph accelerator dimensions and the value of hardware flexibility}.
\newblock \bibinfo{journal}{\emph{IEEE Micro}} \bibinfo{volume}{42}, \bibinfo{number}{4} (\bibinfo{year}{2022}), \bibinfo{pages}{87--96}.
\newblock


\bibitem[Dai et~al\mbox{.}(2019)]%
        {8328836}
\bibfield{author}{\bibinfo{person}{Guohao Dai}, \bibinfo{person}{Tianhao Huang}, \bibinfo{person}{Yuze Chi}, \bibinfo{person}{Jishen Zhao}, \bibinfo{person}{Guangyu Sun}, \bibinfo{person}{Yongpan Liu}, \bibinfo{person}{Yu Wang}, \bibinfo{person}{Yuan Xie}, {and} \bibinfo{person}{Huazhong Yang}.} \bibinfo{year}{2019}\natexlab{}.
\newblock \showarticletitle{GraphH: A Processing-in-Memory Architecture for Large-Scale Graph Processing}.
\newblock \bibinfo{journal}{\emph{IEEE Transactions on Computer-Aided Design of Integrated Circuits and Systems}} \bibinfo{volume}{38}, \bibinfo{number}{4} (\bibinfo{year}{2019}), \bibinfo{pages}{640--653}.
\newblock
\urldef\tempurl%
\url{https://doi.org/10.1109/TCAD.2018.2821565}
\showDOI{\tempurl}


\bibitem[Dally and Towles(2004)]%
        {dally2004principles}
\bibfield{author}{\bibinfo{person}{William~James Dally} {and} \bibinfo{person}{Brian~Patrick Towles}.} \bibinfo{year}{2004}\natexlab{}.
\newblock \bibinfo{booktitle}{\emph{Principles and practices of interconnection networks}}.
\newblock \bibinfo{publisher}{Elsevier}.
\newblock


\bibitem[Everson et~al\mbox{.}(2021)]%
        {everson2021time}
\bibfield{author}{\bibinfo{person}{Luke~R Everson}, \bibinfo{person}{Sachin~S Sapatnekar}, {and} \bibinfo{person}{Chris~H Kim}.} \bibinfo{year}{2021}\natexlab{}.
\newblock \showarticletitle{A Time-Based Intra-Memory Computing Graph Processor Featuring A* Wavefront Expansion and 2-D Gradient Control}.
\newblock \bibinfo{journal}{\emph{IEEE Journal of Solid-State Circuits}} \bibinfo{volume}{56}, \bibinfo{number}{7} (\bibinfo{year}{2021}), \bibinfo{pages}{2281--2290}.
\newblock


\bibitem[Gobieski et~al\mbox{.}(2022)]%
        {riptide}
\bibfield{author}{\bibinfo{person}{Graham Gobieski}, \bibinfo{person}{Souradip Ghosh}, \bibinfo{person}{Marijn Heule}, \bibinfo{person}{Todd Mowry}, \bibinfo{person}{Tony Nowatzki}, \bibinfo{person}{Nathan Beckmann}, {and} \bibinfo{person}{Brandon Lucia}.} \bibinfo{year}{2022}\natexlab{}.
\newblock \showarticletitle{A programmable, energy-minimal dataflow compiler and architecture}. In \bibinfo{booktitle}{\emph{2022 55th IEEE/ACM International Symposium on Microarchitecture (MICRO)}}. IEEE, \bibinfo{pages}{546--564}.
\newblock


\bibitem[Gui et~al\mbox{.}(2019)]%
        {gui2019survey}
\bibfield{author}{\bibinfo{person}{Chuang-Yi Gui}, \bibinfo{person}{Long Zheng}, \bibinfo{person}{Bingsheng He}, \bibinfo{person}{Cheng Liu}, \bibinfo{person}{Xin-Yu Chen}, \bibinfo{person}{Xiao-Fei Liao}, {and} \bibinfo{person}{Hai Jin}.} \bibinfo{year}{2019}\natexlab{}.
\newblock \showarticletitle{A survey on graph processing accelerators: Challenges and opportunities}.
\newblock \bibinfo{journal}{\emph{Journal of Computer Science and Technology}} \bibinfo{volume}{34}, \bibinfo{number}{2} (\bibinfo{year}{2019}), \bibinfo{pages}{339--371}.
\newblock


\bibitem[Guthaus et~al\mbox{.}(2001)]%
        {guthaus2001mibench}
\bibfield{author}{\bibinfo{person}{Matthew~R Guthaus}, \bibinfo{person}{Jeffrey~S Ringenberg}, \bibinfo{person}{Dan Ernst}, \bibinfo{person}{Todd~M Austin}, \bibinfo{person}{Trevor Mudge}, {and} \bibinfo{person}{Richard~B Brown}.} \bibinfo{year}{2001}\natexlab{}.
\newblock \showarticletitle{MiBench: A free, commercially representative embedded benchmark suite}. In \bibinfo{booktitle}{\emph{Proceedings of the fourth annual IEEE international workshop on workload characterization. WWC-4 (Cat. No. 01EX538)}}. IEEE, \bibinfo{pages}{3--14}.
\newblock


\bibitem[Ham et~al\mbox{.}(2016)]%
        {ham2016graphicionado}
\bibfield{author}{\bibinfo{person}{Tae~Jun Ham}, \bibinfo{person}{Lisa Wu}, \bibinfo{person}{Narayanan Sundaram}, \bibinfo{person}{Nadathur Satish}, {and} \bibinfo{person}{Margaret Martonosi}.} \bibinfo{year}{2016}\natexlab{}.
\newblock \showarticletitle{Graphicionado: A high-performance and energy-efficient accelerator for graph analytics}. In \bibinfo{booktitle}{\emph{2016 49th Annual IEEE/ACM International Symposium on Microarchitecture (MICRO)}}. IEEE, \bibinfo{pages}{1--13}.
\newblock


\bibitem[Hamzeh et~al\mbox{.}(2014)]%
        {hamzeh2014branch}
\bibfield{author}{\bibinfo{person}{Mahdi Hamzeh}, \bibinfo{person}{Aviral Shrivastava}, {and} \bibinfo{person}{Sarma Vrudhula}.} \bibinfo{year}{2014}\natexlab{}.
\newblock \showarticletitle{Branch-aware loop mapping on CGRAs}. In \bibinfo{booktitle}{\emph{Proceedings of the 51st Annual Design Automation Conference}}. \bibinfo{pages}{1--6}.
\newblock


\bibitem[Jerger et~al\mbox{.}(2017)]%
        {jerger2017chip}
\bibfield{author}{\bibinfo{person}{Natalie~Enright Jerger}, \bibinfo{person}{Tushar Krishna}, {and} \bibinfo{person}{Li-Shiuan Peh}.} \bibinfo{year}{2017}\natexlab{}.
\newblock \showarticletitle{On-chip networks}.
\newblock \bibinfo{journal}{\emph{Synthesis Lectures on Computer Architecture}} \bibinfo{volume}{12}, \bibinfo{number}{3} (\bibinfo{year}{2017}), \bibinfo{pages}{1--210}.
\newblock


\bibitem[Kapre(2015)]%
        {kapre2015custom}
\bibfield{author}{\bibinfo{person}{Nachiket Kapre}.} \bibinfo{year}{2015}\natexlab{}.
\newblock \showarticletitle{Custom FPGA-based soft-processors for sparse graph acceleration}. In \bibinfo{booktitle}{\emph{2015 IEEE 26th International Conference on Application-specific Systems, Architectures and Processors (ASAP)}}. IEEE, \bibinfo{pages}{9--16}.
\newblock


\bibitem[Karunaratne et~al\mbox{.}(2017)]%
        {karunaratne2017hycube}
\bibfield{author}{\bibinfo{person}{Manupa Karunaratne}, \bibinfo{person}{Aditi~Kulkarni Mohite}, \bibinfo{person}{Tulika Mitra}, {and} \bibinfo{person}{Li-Shiuan Peh}.} \bibinfo{year}{2017}\natexlab{}.
\newblock \showarticletitle{Hycube: A cgra with reconfigurable single-cycle multi-hop interconnect}. In \bibinfo{booktitle}{\emph{Proceedings of the 54th Annual Design Automation Conference 2017}}. \bibinfo{pages}{1--6}.
\newblock


\bibitem[Karunaratne et~al\mbox{.}(2019)]%
        {karunaratne20194d}
\bibfield{author}{\bibinfo{person}{Manupa Karunaratne}, \bibinfo{person}{Dhananjaya Wijerathne}, \bibinfo{person}{Tulika Mitra}, {and} \bibinfo{person}{Li-Shiuan Peh}.} \bibinfo{year}{2019}\natexlab{}.
\newblock \showarticletitle{4D-CGRA: Introducing branch dimension to spatio-temporal application mapping on CGRAs}. In \bibinfo{booktitle}{\emph{2019 IEEE/ACM International Conference on Computer-Aided Design (ICCAD)}}. IEEE, \bibinfo{pages}{1--8}.
\newblock


\bibitem[Lei et~al\mbox{.}(2015)]%
        {lei2015fpga}
\bibfield{author}{\bibinfo{person}{Guoqing Lei}, \bibinfo{person}{Yong Dou}, \bibinfo{person}{Rongchun Li}, {and} \bibinfo{person}{Fei Xia}.} \bibinfo{year}{2015}\natexlab{}.
\newblock \showarticletitle{An FPGA implementation for solving the large single-source-shortest-path problem}.
\newblock \bibinfo{journal}{\emph{IEEE Transactions on Circuits and Systems II: Express Briefs}} \bibinfo{volume}{63}, \bibinfo{number}{5} (\bibinfo{year}{2015}), \bibinfo{pages}{473--477}.
\newblock


\bibitem[Leskovec and Krevl(2014)]%
        {leskovec2014snap}
\bibfield{author}{\bibinfo{person}{Jure Leskovec} {and} \bibinfo{person}{Andrej Krevl}.} \bibinfo{year}{2014}\natexlab{}.
\newblock \bibinfo{title}{SNAP Datasets: Stanford large network dataset collection}.
\newblock
\newblock


\bibitem[Li et~al\mbox{.}(2005)]%
        {li2005trip}
\bibfield{author}{\bibinfo{person}{Feifei Li}, \bibinfo{person}{Dihan Cheng}, \bibinfo{person}{Marios Hadjieleftheriou}, \bibinfo{person}{George Kollios}, {and} \bibinfo{person}{Shang-Hua Teng}.} \bibinfo{year}{2005}\natexlab{}.
\newblock \showarticletitle{On trip planning queries in spatial databases}. In \bibinfo{booktitle}{\emph{International symposium on spatial and temporal databases}}. Springer, \bibinfo{pages}{273--290}.
\newblock


\bibitem[Li et~al\mbox{.}(2022)]%
        {li2022coarse}
\bibfield{author}{\bibinfo{person}{Zhaoying Li}, \bibinfo{person}{D Wijerathne}, {and} \bibinfo{person}{Tulika Mitra}.} \bibinfo{year}{2022}\natexlab{}.
\newblock \showarticletitle{Coarse Grained Reconfigurable Array CGRA}.
\newblock \bibinfo{journal}{\emph{Book Chapter in Springer Handbook of Computer Architecture}} (\bibinfo{year}{2022}).
\newblock


\bibitem[Liu et~al\mbox{.}(2019)]%
        {liu2019survey}
\bibfield{author}{\bibinfo{person}{Leibo Liu}, \bibinfo{person}{Jianfeng Zhu}, \bibinfo{person}{Zhaoshi Li}, \bibinfo{person}{Yanan Lu}, \bibinfo{person}{Yangdong Deng}, \bibinfo{person}{Jie Han}, \bibinfo{person}{Shouyi Yin}, {and} \bibinfo{person}{Shaojun Wei}.} \bibinfo{year}{2019}\natexlab{}.
\newblock \showarticletitle{A survey of coarse-grained reconfigurable architecture and design: Taxonomy, challenges, and applications}.
\newblock \bibinfo{journal}{\emph{ACM Computing Surveys (CSUR)}} \bibinfo{volume}{52}, \bibinfo{number}{6} (\bibinfo{year}{2019}), \bibinfo{pages}{1--39}.
\newblock


\bibitem[Liu et~al\mbox{.}(2022)]%
        {liu2022overgen}
\bibfield{author}{\bibinfo{person}{Sihao Liu}, \bibinfo{person}{Jian Weng}, \bibinfo{person}{Dylan Kupsh}, \bibinfo{person}{Atefeh Sohrabizadeh}, \bibinfo{person}{Zhengrong Wang}, \bibinfo{person}{Licheng Guo}, \bibinfo{person}{Jiuyang Liu}, \bibinfo{person}{Maxim Zhulin}, \bibinfo{person}{Rishabh Mani}, \bibinfo{person}{Lucheng Zhang}, {et~al\mbox{.}}} \bibinfo{year}{2022}\natexlab{}.
\newblock \showarticletitle{OverGen: Improving FPGA usability through domain-specific overlay generation}. In \bibinfo{booktitle}{\emph{2022 55th IEEE/ACM International Symposium on Microarchitecture (MICRO)}}. IEEE, \bibinfo{pages}{35--56}.
\newblock


\bibitem[Low et~al\mbox{.}(2014)]%
        {low2014graphlab}
\bibfield{author}{\bibinfo{person}{Yucheng Low}, \bibinfo{person}{Joseph~E Gonzalez}, \bibinfo{person}{Aapo Kyrola}, \bibinfo{person}{Danny Bickson}, \bibinfo{person}{Carlos~E Guestrin}, {and} \bibinfo{person}{Joseph Hellerstein}.} \bibinfo{year}{2014}\natexlab{}.
\newblock \showarticletitle{Graphlab: A new framework for parallel machine learning}.
\newblock \bibinfo{journal}{\emph{arXiv preprint arXiv:1408.2041}} (\bibinfo{year}{2014}).
\newblock


\bibitem[Martin(2022)]%
        {martin2022twenty}
\bibfield{author}{\bibinfo{person}{Kevin~JM Martin}.} \bibinfo{year}{2022}\natexlab{}.
\newblock \showarticletitle{Twenty Years of Automated Methods for Mapping Applications on CGRA}. In \bibinfo{booktitle}{\emph{2022 IEEE International Parallel and Distributed Processing Symposium Workshops (IPDPSW)}}. IEEE, \bibinfo{pages}{679--686}.
\newblock


\bibitem[McCune et~al\mbox{.}(2015)]%
        {mccune2015thinking}
\bibfield{author}{\bibinfo{person}{Robert~Ryan McCune}, \bibinfo{person}{Tim Weninger}, {and} \bibinfo{person}{Greg Madey}.} \bibinfo{year}{2015}\natexlab{}.
\newblock \showarticletitle{Thinking like a vertex: a survey of vertex-centric frameworks for large-scale distributed graph processing}.
\newblock \bibinfo{journal}{\emph{ACM Computing Surveys (CSUR)}} \bibinfo{volume}{48}, \bibinfo{number}{2} (\bibinfo{year}{2015}), \bibinfo{pages}{1--39}.
\newblock


\bibitem[Mishra et~al\mbox{.}(2006)]%
        {mishra2006tartan}
\bibfield{author}{\bibinfo{person}{Mahim Mishra}, \bibinfo{person}{Timothy~J Callahan}, \bibinfo{person}{Tiberiu Chelcea}, \bibinfo{person}{Girish Venkataramani}, \bibinfo{person}{Seth~C Goldstein}, {and} \bibinfo{person}{Mihai Budiu}.} \bibinfo{year}{2006}\natexlab{}.
\newblock \showarticletitle{Tartan: evaluating spatial computation for whole program execution}.
\newblock \bibinfo{journal}{\emph{ACM SIGARCH Computer Architecture News}} \bibinfo{volume}{34}, \bibinfo{number}{5} (\bibinfo{year}{2006}), \bibinfo{pages}{163--174}.
\newblock


\bibitem[Nguyen and Sanchez(2021)]%
        {nguyen2021fifer}
\bibfield{author}{\bibinfo{person}{Quan~M Nguyen} {and} \bibinfo{person}{Daniel Sanchez}.} \bibinfo{year}{2021}\natexlab{}.
\newblock \showarticletitle{Fifer: Practical acceleration of irregular applications on reconfigurable architectures}. In \bibinfo{booktitle}{\emph{MICRO-54: 54th Annual IEEE/ACM International Symposium on Microarchitecture}}. \bibinfo{pages}{1064--1077}.
\newblock


\bibitem[Nowatzki et~al\mbox{.}(2017)]%
        {nowatzki2017stream}
\bibfield{author}{\bibinfo{person}{Tony Nowatzki}, \bibinfo{person}{Vinay Gangadhar}, \bibinfo{person}{Newsha Ardalani}, {and} \bibinfo{person}{Karthikeyan Sankaralingam}.} \bibinfo{year}{2017}\natexlab{}.
\newblock \showarticletitle{Stream-dataflow acceleration}. In \bibinfo{booktitle}{\emph{2017 ACM/IEEE 44th Annual International Symposium on Computer Architecture (ISCA)}}. IEEE, \bibinfo{pages}{416--429}.
\newblock


\bibitem[Ozdal et~al\mbox{.}(2016)]%
        {ozdal2016energy}
\bibfield{author}{\bibinfo{person}{Muhammet~Mustafa Ozdal}, \bibinfo{person}{Serif Yesil}, \bibinfo{person}{Taemin Kim}, \bibinfo{person}{Andrey Ayupov}, \bibinfo{person}{John Greth}, \bibinfo{person}{Steven Burns}, {and} \bibinfo{person}{Ozcan Ozturk}.} \bibinfo{year}{2016}\natexlab{}.
\newblock \showarticletitle{Energy efficient architecture for graph analytics accelerators}.
\newblock \bibinfo{journal}{\emph{ACM SIGARCH Computer Architecture News}} \bibinfo{volume}{44}, \bibinfo{number}{3} (\bibinfo{year}{2016}), \bibinfo{pages}{166--177}.
\newblock


\bibitem[Pawlowski(2011)]%
        {pawlowski2011hybrid}
\bibfield{author}{\bibinfo{person}{J~Thomas Pawlowski}.} \bibinfo{year}{2011}\natexlab{}.
\newblock \showarticletitle{Hybrid memory cube (HMC)}. In \bibinfo{booktitle}{\emph{2011 IEEE Hot chips 23 symposium (HCS)}}. IEEE, \bibinfo{pages}{1--24}.
\newblock


\bibitem[Podobas et~al\mbox{.}(2020)]%
        {podobas2020survey}
\bibfield{author}{\bibinfo{person}{Artur Podobas}, \bibinfo{person}{Kentaro Sano}, {and} \bibinfo{person}{Satoshi Matsuoka}.} \bibinfo{year}{2020}\natexlab{}.
\newblock \showarticletitle{A survey on coarse-grained reconfigurable architectures from a performance perspective}.
\newblock \bibinfo{journal}{\emph{IEEE Access}}  \bibinfo{volume}{8} (\bibinfo{year}{2020}), \bibinfo{pages}{146719--146743}.
\newblock


\bibitem[Rahman et~al\mbox{.}(2020)]%
        {rahman2020graphpulse}
\bibfield{author}{\bibinfo{person}{Shafiur Rahman}, \bibinfo{person}{Nael Abu-Ghazaleh}, {and} \bibinfo{person}{Rajiv Gupta}.} \bibinfo{year}{2020}\natexlab{}.
\newblock \showarticletitle{Graphpulse: An event-driven hardware accelerator for asynchronous graph processing}. In \bibinfo{booktitle}{\emph{2020 53rd Annual IEEE/ACM International Symposium on Microarchitecture (MICRO)}}. IEEE, \bibinfo{pages}{908--921}.
\newblock


\bibitem[Reddy et~al\mbox{.}(1977)]%
        {reddy1977speech}
\bibfield{author}{\bibinfo{person}{D~Raj Reddy} {et~al\mbox{.}}} \bibinfo{year}{1977}\natexlab{}.
\newblock \showarticletitle{Speech understanding systems: A summary of results of the five-year research effort}.
\newblock \bibinfo{journal}{\emph{Department of Computer Science. Camegie-Mell University, Pittsburgh, PA}}  \bibinfo{volume}{17} (\bibinfo{year}{1977}), \bibinfo{pages}{138}.
\newblock


\bibitem[Song et~al\mbox{.}(2018)]%
        {song2018graphr}
\bibfield{author}{\bibinfo{person}{Linghao Song}, \bibinfo{person}{Youwei Zhuo}, \bibinfo{person}{Xuehai Qian}, \bibinfo{person}{Hai Li}, {and} \bibinfo{person}{Yiran Chen}.} \bibinfo{year}{2018}\natexlab{}.
\newblock \showarticletitle{GraphR: Accelerating graph processing using ReRAM}. In \bibinfo{booktitle}{\emph{2018 IEEE International Symposium on High Performance Computer Architecture (HPCA)}}. IEEE, \bibinfo{pages}{531--543}.
\newblock


\bibitem[Stevens et~al\mbox{.}(2021)]%
        {stevens2021gnnerator}
\bibfield{author}{\bibinfo{person}{Jacob~R Stevens}, \bibinfo{person}{Dipankar Das}, \bibinfo{person}{Sasikanth Avancha}, \bibinfo{person}{Bharat Kaul}, {and} \bibinfo{person}{Anand Raghunathan}.} \bibinfo{year}{2021}\natexlab{}.
\newblock \showarticletitle{GNNerator: A hardware/software framework for accelerating graph neural networks}. In \bibinfo{booktitle}{\emph{2021 58th ACM/IEEE Design Automation Conference (DAC)}}. IEEE, \bibinfo{pages}{955--960}.
\newblock


\bibitem[Swanson et~al\mbox{.}(2003)]%
        {swanson2003wavescalar}
\bibfield{author}{\bibinfo{person}{Steven Swanson}, \bibinfo{person}{Ken Michelson}, \bibinfo{person}{Andrew Schwerin}, {and} \bibinfo{person}{Mark Oskin}.} \bibinfo{year}{2003}\natexlab{}.
\newblock \showarticletitle{WaveScalar}. In \bibinfo{booktitle}{\emph{Proceedings. 36th Annual IEEE/ACM International Symposium on Microarchitecture, 2003. MICRO-36.}} IEEE, \bibinfo{pages}{291--302}.
\newblock


\bibitem[Tan et~al\mbox{.}(2021a)]%
        {tan2021opencgra}
\bibfield{author}{\bibinfo{person}{Cheng Tan}, \bibinfo{person}{Nicolas~Bohm Agostini}, \bibinfo{person}{Jeff Zhang}, \bibinfo{person}{Marco Minutoli}, \bibinfo{person}{Vito~Giovanni Castellana}, \bibinfo{person}{Chenhao Xie}, \bibinfo{person}{Tong Geng}, \bibinfo{person}{Ang Li}, \bibinfo{person}{Kevin Barker}, {and} \bibinfo{person}{Antonino Tumeo}.} \bibinfo{year}{2021}\natexlab{a}.
\newblock \showarticletitle{Opencgra: Democratizing coarse-grained reconfigurable arrays}. In \bibinfo{booktitle}{\emph{2021 IEEE 32nd International Conference on Application-specific Systems, Architectures and Processors (ASAP)}}. IEEE, \bibinfo{pages}{149--155}.
\newblock


\bibitem[Tan et~al\mbox{.}(2022)]%
        {tan2022asap}
\bibfield{author}{\bibinfo{person}{Cheng Tan}, \bibinfo{person}{Thierry Tambe}, \bibinfo{person}{Jeff Zhang}, \bibinfo{person}{Bo Fang}, \bibinfo{person}{Tong Geng}, \bibinfo{person}{Gu-Yeon Wei}, \bibinfo{person}{David Brooks}, \bibinfo{person}{Antonino Tumeo}, \bibinfo{person}{Ganesh Gopalakrishnan}, {and} \bibinfo{person}{Ang Li}.} \bibinfo{year}{2022}\natexlab{}.
\newblock \showarticletitle{ASAP: automatic synthesis of area-efficient and precision-aware CGRAs}. In \bibinfo{booktitle}{\emph{Proceedings of the 36th ACM International Conference on Supercomputing}}. \bibinfo{pages}{1--13}.
\newblock


\bibitem[Tan et~al\mbox{.}(2020)]%
        {tan2020opencgra}
\bibfield{author}{\bibinfo{person}{Cheng Tan}, \bibinfo{person}{Chenhao Xie}, \bibinfo{person}{Ang Li}, \bibinfo{person}{Kevin~J Barker}, {and} \bibinfo{person}{Antonino Tumeo}.} \bibinfo{year}{2020}\natexlab{}.
\newblock \showarticletitle{OpenCGRA: An open-source unified framework for modeling, testing, and evaluating CGRAs}. In \bibinfo{booktitle}{\emph{2020 IEEE 38th International Conference on Computer Design (ICCD)}}. IEEE, \bibinfo{pages}{381--388}.
\newblock


\bibitem[Tan et~al\mbox{.}(2021b)]%
        {tan2021aurora}
\bibfield{author}{\bibinfo{person}{Cheng Tan}, \bibinfo{person}{Chenhao Xie}, \bibinfo{person}{Ang Li}, \bibinfo{person}{Kevin~J Barker}, {and} \bibinfo{person}{Antonino Tumeo}.} \bibinfo{year}{2021}\natexlab{b}.
\newblock \showarticletitle{Aurora: Automated refinement of coarse-grained reconfigurable accelerators}. In \bibinfo{booktitle}{\emph{2021 Design, Automation \& Test in Europe Conference \& Exhibition (DATE)}}. IEEE, \bibinfo{pages}{1388--1393}.
\newblock


\bibitem[Voitsechov and Etsion(2014)]%
        {voitsechov2014single}
\bibfield{author}{\bibinfo{person}{Dani Voitsechov} {and} \bibinfo{person}{Yoav Etsion}.} \bibinfo{year}{2014}\natexlab{}.
\newblock \showarticletitle{Single-graph multiple flows: Energy efficient design alternative for GPGPUs}.
\newblock \bibinfo{journal}{\emph{ACM SIGARCH computer architecture news}} \bibinfo{volume}{42}, \bibinfo{number}{3} (\bibinfo{year}{2014}), \bibinfo{pages}{205--216}.
\newblock


\bibitem[Waingold et~al\mbox{.}(1997)]%
        {waingold1997baring}
\bibfield{author}{\bibinfo{person}{Elliot Waingold}, \bibinfo{person}{Michael Taylor}, \bibinfo{person}{Devabhaktuni Srikrishna}, \bibinfo{person}{Vivek Sarkar}, \bibinfo{person}{Walter Lee}, \bibinfo{person}{Victor Lee}, \bibinfo{person}{Jang Kim}, \bibinfo{person}{Matthew Frank}, \bibinfo{person}{Peter Finch}, \bibinfo{person}{Rajeev Barua}, {et~al\mbox{.}}} \bibinfo{year}{1997}\natexlab{}.
\newblock \showarticletitle{Baring it all to software: Raw machines}.
\newblock \bibinfo{journal}{\emph{Computer}} \bibinfo{volume}{30}, \bibinfo{number}{9} (\bibinfo{year}{1997}), \bibinfo{pages}{86--93}.
\newblock


\bibitem[Wang et~al\mbox{.}(2019)]%
        {wang2019hycube}
\bibfield{author}{\bibinfo{person}{Bo Wang}, \bibinfo{person}{Manupa Karunarathne}, \bibinfo{person}{Aditi Kulkarni}, \bibinfo{person}{Tulika Mitra}, {and} \bibinfo{person}{Li-Shiuan Peh}.} \bibinfo{year}{2019}\natexlab{}.
\newblock \showarticletitle{Hycube: A 0.9 v 26.4 mops/mw, 290 pj/op, power efficient accelerator for iot applications}. In \bibinfo{booktitle}{\emph{2019 IEEE Asian Solid-State Circuits Conference (A-SSCC)}}. IEEE, \bibinfo{pages}{133--136}.
\newblock


\bibitem[Weng et~al\mbox{.}(2020a)]%
        {weng2020dsagen}
\bibfield{author}{\bibinfo{person}{Jian Weng}, \bibinfo{person}{Sihao Liu}, \bibinfo{person}{Vidushi Dadu}, \bibinfo{person}{Zhengrong Wang}, \bibinfo{person}{Preyas Shah}, {and} \bibinfo{person}{Tony Nowatzki}.} \bibinfo{year}{2020}\natexlab{a}.
\newblock \showarticletitle{Dsagen: Synthesizing programmable spatial accelerators}. In \bibinfo{booktitle}{\emph{2020 ACM/IEEE 47th Annual International Symposium on Computer Architecture (ISCA)}}. IEEE, \bibinfo{pages}{268--281}.
\newblock


\bibitem[Weng et~al\mbox{.}(2020b)]%
        {weng2020hybrid}
\bibfield{author}{\bibinfo{person}{Jian Weng}, \bibinfo{person}{Sihao Liu}, \bibinfo{person}{Zhengrong Wang}, \bibinfo{person}{Vidushi Dadu}, {and} \bibinfo{person}{Tony Nowatzki}.} \bibinfo{year}{2020}\natexlab{b}.
\newblock \showarticletitle{A hybrid systolic-dataflow architecture for inductive matrix algorithms}. In \bibinfo{booktitle}{\emph{2020 IEEE International Symposium on High Performance Computer Architecture (HPCA)}}. IEEE, \bibinfo{pages}{703--716}.
\newblock


\bibitem[Wijerathne et~al\mbox{.}(2019)]%
        {wijerathne2019cascade}
\bibfield{author}{\bibinfo{person}{Dhananjaya Wijerathne}, \bibinfo{person}{Zhaoying Li}, \bibinfo{person}{Manupa Karunarathne}, \bibinfo{person}{Anuj Pathania}, {and} \bibinfo{person}{Tulika Mitra}.} \bibinfo{year}{2019}\natexlab{}.
\newblock \showarticletitle{Cascade: High throughput data streaming via decoupled access-execute cgra}.
\newblock \bibinfo{journal}{\emph{ACM Transactions on Embedded Computing Systems (TECS)}} \bibinfo{volume}{18}, \bibinfo{number}{5s} (\bibinfo{year}{2019}), \bibinfo{pages}{1--26}.
\newblock


\bibitem[Wijerathne et~al\mbox{.}(2022)]%
        {wijerathne2022morpher}
\bibfield{author}{\bibinfo{person}{Dhananjaya Wijerathne}, \bibinfo{person}{Zhaoying Li}, \bibinfo{person}{Manupa Karunaratne}, \bibinfo{person}{Li-Shiuan Peh}, {and} \bibinfo{person}{Tulika Mitra}.} \bibinfo{year}{2022}\natexlab{}.
\newblock \showarticletitle{Morpher: An Open-Source Integrated Compilation and Simulation Framework for CGRA}. In \bibinfo{booktitle}{\emph{Fifth Workshop on Open-Source EDA Technology (WOSET)}}.
\newblock


\bibitem[Wong et~al\mbox{.}(2012)]%
        {wong2012metal}
\bibfield{author}{\bibinfo{person}{H-S~Philip Wong}, \bibinfo{person}{Heng-Yuan Lee}, \bibinfo{person}{Shimeng Yu}, \bibinfo{person}{Yu-Sheng Chen}, \bibinfo{person}{Yi Wu}, \bibinfo{person}{Pang-Shiu Chen}, \bibinfo{person}{Byoungil Lee}, \bibinfo{person}{Frederick~T Chen}, {and} \bibinfo{person}{Ming-Jinn Tsai}.} \bibinfo{year}{2012}\natexlab{}.
\newblock \showarticletitle{Metal--oxide RRAM}.
\newblock \bibinfo{journal}{\emph{Proc. IEEE}} \bibinfo{volume}{100}, \bibinfo{number}{6} (\bibinfo{year}{2012}), \bibinfo{pages}{1951--1970}.
\newblock


\bibitem[Yao et~al\mbox{.}(2022)]%
        {yao2022scalagraph}
\bibfield{author}{\bibinfo{person}{Pengcheng Yao}, \bibinfo{person}{Long Zheng}, \bibinfo{person}{Yu Huang}, \bibinfo{person}{Qinggang Wang}, \bibinfo{person}{Chuangyi Gui}, \bibinfo{person}{Zhen Zeng}, \bibinfo{person}{Xiaofei Liao}, \bibinfo{person}{Hai Jin}, {and} \bibinfo{person}{Jingling Xue}.} \bibinfo{year}{2022}\natexlab{}.
\newblock \showarticletitle{Scalagraph: A scalable accelerator for massively parallel graph processing}. In \bibinfo{booktitle}{\emph{2022 IEEE International Symposium on High-Performance Computer Architecture (HPCA)}}. IEEE, \bibinfo{pages}{199--212}.
\newblock


\bibitem[Yuan et~al\mbox{.}(2021)]%
        {yuan2021dynamic}
\bibfield{author}{\bibinfo{person}{Baofen Yuan}, \bibinfo{person}{Jianfeng Zhu}, \bibinfo{person}{Xingchen Man}, \bibinfo{person}{Zijiao Ma}, \bibinfo{person}{Shouyi Yin}, \bibinfo{person}{Shaojun Wei}, {and} \bibinfo{person}{Leibo Liu}.} \bibinfo{year}{2021}\natexlab{}.
\newblock \showarticletitle{Dynamic-II Pipeline: Compiling Loops with Irregular Branches on Static-Scheduling CGRA}.
\newblock \bibinfo{journal}{\emph{IEEE Transactions on Computer-Aided Design of Integrated Circuits and Systems}} (\bibinfo{year}{2021}).
\newblock


\bibitem[Zhang et~al\mbox{.}(2018)]%
        {zhang2018graphp}
\bibfield{author}{\bibinfo{person}{Mingxing Zhang}, \bibinfo{person}{Youwei Zhuo}, \bibinfo{person}{Chao Wang}, \bibinfo{person}{Mingyu Gao}, \bibinfo{person}{Yongwei Wu}, \bibinfo{person}{Kang Chen}, \bibinfo{person}{Christos Kozyrakis}, {and} \bibinfo{person}{Xuehai Qian}.} \bibinfo{year}{2018}\natexlab{}.
\newblock \showarticletitle{GraphP: Reducing communication for PIM-based graph processing with efficient data partition}. In \bibinfo{booktitle}{\emph{2018 IEEE International Symposium on High Performance Computer Architecture (HPCA)}}. IEEE, \bibinfo{pages}{544--557}.
\newblock


\bibitem[Zhang et~al\mbox{.}(2021)]%
        {zhang2021depgraph}
\bibfield{author}{\bibinfo{person}{Yu Zhang}, \bibinfo{person}{Xiaofei Liao}, \bibinfo{person}{Hai Jin}, \bibinfo{person}{Ligang He}, \bibinfo{person}{Bingsheng He}, \bibinfo{person}{Haikun Liu}, {and} \bibinfo{person}{Lin Gu}.} \bibinfo{year}{2021}\natexlab{}.
\newblock \showarticletitle{Depgraph: A dependency-driven accelerator for efficient iterative graph processing}. In \bibinfo{booktitle}{\emph{2021 IEEE International Symposium on High-Performance Computer Architecture (HPCA)}}. IEEE, \bibinfo{pages}{371--384}.
\newblock


\bibitem[Zhou et~al\mbox{.}(2017)]%
        {zhou2017tunao}
\bibfield{author}{\bibinfo{person}{Jinhong Zhou}, \bibinfo{person}{Shaoli Liu}, \bibinfo{person}{Qi Guo}, \bibinfo{person}{Xuda Zhou}, \bibinfo{person}{Tian Zhi}, \bibinfo{person}{Daofu Liu}, \bibinfo{person}{Chao Wang}, \bibinfo{person}{Xuehai Zhou}, \bibinfo{person}{Yunji Chen}, {and} \bibinfo{person}{Tianshi Chen}.} \bibinfo{year}{2017}\natexlab{}.
\newblock \showarticletitle{Tunao: A high-performance and energy-efficient reconfigurable accelerator for graph processing}. In \bibinfo{booktitle}{\emph{2017 17th IEEE/ACM International Symposium on Cluster, Cloud and Grid Computing (CCGRID)}}. IEEE, \bibinfo{pages}{731--734}.
\newblock


\bibitem[Zhou et~al\mbox{.}(2018)]%
        {zhou2018fpga}
\bibfield{author}{\bibinfo{person}{Shijie Zhou}, \bibinfo{person}{Rajgopal Kannan}, \bibinfo{person}{Hanqing Zeng}, {and} \bibinfo{person}{Viktor~K Prasanna}.} \bibinfo{year}{2018}\natexlab{}.
\newblock \showarticletitle{An FPGA framework for edge-centric graph processing}. In \bibinfo{booktitle}{\emph{Proceedings of the 15th ACM International Conference on Computing Frontiers}}. \bibinfo{pages}{69--77}.
\newblock


\bibitem[Zhuo et~al\mbox{.}(2019)]%
        {10.1145/3352460.3358256}
\bibfield{author}{\bibinfo{person}{Youwei Zhuo}, \bibinfo{person}{Chao Wang}, \bibinfo{person}{Mingxing Zhang}, \bibinfo{person}{Rui Wang}, \bibinfo{person}{Dimin Niu}, \bibinfo{person}{Yanzhi Wang}, {and} \bibinfo{person}{Xuehai Qian}.} \bibinfo{year}{2019}\natexlab{}.
\newblock \showarticletitle{GraphQ: Scalable PIM-Based Graph Processing}. In \bibinfo{booktitle}{\emph{Proceedings of the 52nd Annual IEEE/ACM International Symposium on Microarchitecture}} (Columbus, OH, USA) \emph{(\bibinfo{series}{MICRO '52})}. \bibinfo{publisher}{Association for Computing Machinery}, \bibinfo{address}{New York, NY, USA}, \bibinfo{pages}{712–725}.
\newblock
\showISBNx{9781450369381}
\urldef\tempurl%
\url{https://doi.org/10.1145/3352460.3358256}
\showDOI{\tempurl}


\end{thebibliography}

\end{document}